\documentclass[a4paper,fleqn]{cas-sc}

\usepackage[numbers,longnamesfirst]{natbib}
\usepackage[english]{babel}
\usepackage{booktabs} %
\usepackage{fancyvrb}
\usepackage{listings}
\usepackage{color}
\usepackage{balance}
\usepackage{fancybox}
\usepackage{xspace}
\usepackage{subcaption}
\usepackage{multirow}
\usepackage{url}
\usepackage{array}
\usepackage{graphicx}
\usepackage{bbding}
\usepackage{wasysym}
\usepackage{amssymb}
\usepackage{pdflscape}
\usepackage{tabularx,colortbl}
\usepackage{dirtree}
\usepackage{rotating}
\usepackage{fancyvrb}
\usepackage{comment}
\usepackage{soul}
\usepackage{algorithm}
\usepackage[noend]{algpseudocode}

\usepackage{enumitem}

\usepackage{colortbl}
\usepackage{tabu}
\usepackage{xcolor}
\usepackage[skins]{tcolorbox}
\usepackage[noend]{algpseudocode}
\usepackage{cleveref}
\usepackage{footmisc}

\usepackage{float}

\floatstyle{ruled}
\newfloat{program}{thp}{lop}
\floatname{program}{Program}

\usepackage{array, booktabs, boldline} %

\newcommand{\conclusion}[1]{\begin{center}\begin{tcolorbox}[skin=widget, left=2mm,right=2mm,top=2mm,bottom=2mm,boxrule=0.3mm,arc=0mm,coltitle=black,colframe=black!99!white,colback=white!99!gray,width=(\linewidth),before=\hfill,after=\hfill]#1\end{tcolorbox}\end{center}}

\newlength\MAX  

\definecolor{snsblue}{RGB}{86,125,185}
\newcommand\sbar[2]{{\color{snsblue}\rule{\dimexpr 1cm * #1 / #2}{6pt}}}

\newcommand{\npm}{npm\xspace}
\newcommand{\npms}{\textit{npms}\xspace}
\newcommand{\used}{highly-selected\xspace}

\newcommand{\unused}{not highly-selected\xspace}

\newcommand{\term}[1]{\textit{#1}}

\usepackage[
  disable,
  colorinlistoftodos
]{todonotes}
\usepackage{verbatim}

\makeatletter
\if@todonotes@disabled
\else
\paperwidth=\dimexpr \paperwidth + 5cm\relax
\evensidemargin=\dimexpr\evensidemargin + 3.5cm\relax
\marginparwidth=\dimexpr \marginparwidth + 3.5cm\relax
\fi
\makeatother

\definecolor{SkyBlue}{RGB}{79,189,204}
\definecolor{YellowGreen}{RGB}{143,198,109}
\definecolor{Apricot}{RGB}{242,179,126}
\definecolor{Goldenrod}{RGB}{254,220,77}

\newcommand*\fullcaption[2][]{\caption[#1]{#1\xspace#2}}

\definecolor{Gray}{gray}{0.9}
\newcolumntype{g}{>{\columncolor{Gray}}l}

\definecolor{dkgreen}{rgb}{0,0.6,0}
\definecolor{gray}{rgb}{0.5,0.5,0.5}
\definecolor{mauve}{rgb}{0.58,0,0.82}

\def\tsc#1{\csdef{#1}{\textsc{\lowercase{#1}}\xspace}}
\tsc{WGM}
\tsc{QE}

\begin{document}

	\shorttitle{Characteristics of \used packages}
	
	\shortauthors{Mujahid et~al.}
	
	\title [mode = title]{What are the characteristics of \used packages? A case study on the \npm ecosystem}

	\author[1]{Suhaib Mujahid}[orcid=0000-0003-2738-1387]
	\ead{suhaib.mujahid@concordia.ca}
	\address[1]{Data-driven Analysis of Software (DAS) Lab, Concordia University, Montreal, Canada}

\author[2]{Rabe Abdalkareem}[orcid=0000-0001-9914-5434]
\ead{rabe.abdalkareem@carleton.ca}
\address[2]{School of Computer Science, Carleton University, Ottawa, Canada}

	\author[1]{Emad Shihab}[orcid=0000-0003-1285-9878]
	\ead{emad.shihab@concordia.ca}

  \begin{abstract}
With the popularity of software ecosystems, the number of open source components (known as packages) has grown rapidly. Identifying high-quality and well-maintained packages from a large pool of packages to depend on is a basic and important problem, as it is beneficial for various applications, such as package recommendation and package search. However, no systematic and comprehensive work focuses on addressing this problem except in online discussions or informal literature and interviews. To fill this gap, in this paper, we conducted a mixed qualitative and quantitative analysis to understand how developers identify and select relevant open source packages. In particular, we started by surveying 118 JavaScript developers from the \npm ecosystem to qualitatively understand the factors that make a package to be \used within the \npm ecosystem. The survey results showed that JavaScript developers believe that \used packages are well-documented, receive a high number of stars on GitHub, have a large number of downloads, and do not suffer from vulnerabilities. Then, we conducted an experiment to quantitatively validate the developers' perception of the factors that make a \used package. In this analysis, we collected and mined historical data from 2,527 packages divided into \used and \unused packages. For each package in the dataset, we collected quantitative data to present the factors studied in the developers' survey. Next, we used regression analysis to quantitatively investigate which of the studied factors are the most important. Our regression analysis complements our survey results about \used packages. In particular, the results showed that \used packages tend to be correlated by the number of downloads, stars, and how large the package's readme file is.

  \end{abstract}


\maketitle

\section{Introduction}
\label{sec:survey:introduction}
In recent years, the proliferation of software ecosystems has led to a vast and rapid growth of the number of open-source packages.\footnote{In this paper, we use the term package referring to open source components published on software ecosystems.}
For example, as of January~2022, there were over a~million packages available on the registry of the Node Package Manager (\npm), one of the largest software ecosystems.
Furthermore, the number of packages grew by around 60\% between January 2019 and January~2022~\cite{DeBill_Modulecounts:online}.
With the massive number of packages out there, finding the right package to use can be challenging, considering that many packages provide similar functionalities.
However, there are some packages that stand out and experience high interest from developers.
We believe that understanding the characteristics of these \used packages is very important since it helps developers answer the essential question: which packages a developer should select among many existing options.
In addition, such understanding can help to improve the performance of package recommendation systems~\cite{zheng2011FSE,Mora_PROMISE2018,nodejsHo83online,Semeteys_OSB2008} and enhance the user experience of package search engines~\cite{Abdellatif_IST2020,nodejsHo83online,npms13online}. Furthermore, for package developers, understanding the characteristics of \used packages can be helpful for various purposes, such as improving their packages to meet the requirements and experiences that users look for, acquiring the community's attention, and eventually increasing the package usage and overall reputation.
Those potential implications of understanding the factors of \used motivate use to study them.

\begin{figure}[]
	\centering
	\includegraphics[trim=2.8in 3.5in 2.8in 3.5in, clip,width=.7\linewidth]{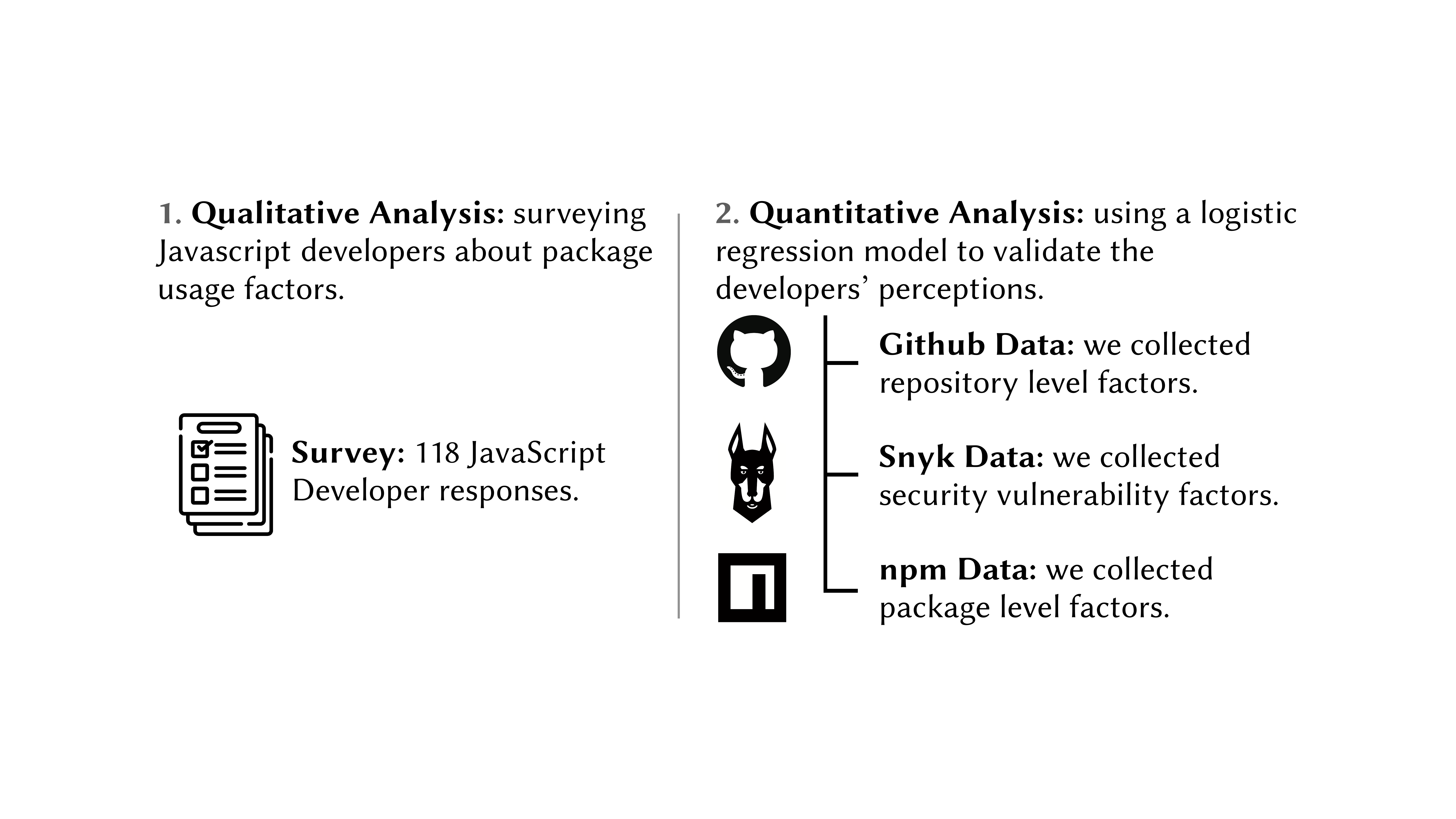}
	\caption{An overview of our study design.}
	\label{fig:survey_approach}
\end{figure}

Previous studies examined different aspects of packages published in software ecosystems, such as centrality and popularity~\cite{Abdalkareem_EMSE2020,Abdellatif_IST2020,Larios2020ACM,qiu2018understanding,Mujahid_arXiv2021}, and a few examined the selection factors of relevant packages~\cite{Larios2020ACM,Jadhav_IST2009}.
The main limitation of prior works is that they were based on a purely quantitative analysis of popular packages or only interviewing developers in a specific industrial context.
That said, understanding the characteristics of \used packages is still the subject of much discussion and refinement.
This is because several facts include personality aspects and examining different data modalities from several sources, and a developer is typically a familiar user of a specific package.
Thus, in this paper, we divided the study into two parts - qualitative and quantitative~\cite{john2000designing}.
\Cref{fig:survey_approach} shows an overview of our study design.
In the first part (referred from now on as \term{qualitative analysis}), we conducted a user study survey that involves {118} JavaScript developers.
We asked our survey participants to fill in a form composed of 17 statements about factors they use when selecting \npm packages.
Then, we qualitatively analyzed the developers answers to the 17 questions using descriptive statistics.

In order to provide validation to the findings of the qualitative analysis, we conducted \term{quantitative analysis} on a set of {2,527} \npm packages grouped into \used and \unused packages.
Similar to prior work~\cite{BavotaTSE2015,lee2020empirical,Tian2015ICSME}, we estimated the \used packages based on the number of directly dependent packages (i.e., clients packages) within the \npm ecosystem.
Then, we mined and analyzed the selected packages and collected quantitative data to present the factors studied in our survey.
Next, we used regression analysis to quantitatively investigate which of the studied factors are the most important for developers to select a package to use.

The survey results showed that JavaScript developers believe that when selecting a package to use, they look for packages that are: well-documented, receive a high number of stars on GitHub, have a large number of downloads, and do not suffer from security vulnerabilities.
Moreover, our regression analysis complemented our survey results about \used packages. For example, our developers' survey and regression analysis revealed that developers select packages with a high number of stars and downloads.
Also, it described the differences between the developers' perceptions about \used packages and their characteristics.
In general, our work makes the following key contributions:

\begin{itemize}%
	\item We performed a mixed qualitative and quantitative analysis to investigate the characteristics of \used packages on the \npm ecosystem.
	      We presented our results from surveying 118 JavaScript developers and validated the survey results through a quantitative analysis of 2,427 \npm packages.

	\item We identified the most important factors that packages' users should consider when selecting an \npm package to use in their projects.
	\item We also provided practical implications for packages' maintainers, the \npm ecosystem's maintainers, and researchers and outline future research avenues.

\end{itemize}

The remainder of this paper is structured as follows.
\Cref{sec:survey:qualitative_analysis} describes the study design and presents the results of the qualitative analysis.
\Cref{sec:survey:quantitative_analysis} describes the study design and presents the results of the quantitative analysis.
We discuss the implications of our study in \Cref{sec:survey:discussion}.
In \Cref{sec:survey:related_work}, we present the work that is related to our study.
We discuss the threats that may affect the validity of the results in \Cref{sec:survey:threats_to_validity}.
Finally, \Cref{sec:survey:conclusion} concludes our work.

\section{Qualitative Analysis}
\label{sec:survey:qualitative_analysis}

This analysis aims to survey JavaScript developers to understand the characteristics of packages that JavaScript developers look for when selecting an \npm package to use.
In this study, we surveyed 118 JavaScript developers.

\subsection{Study Design}
This section presents our survey design, participant recruitment, and data analysis methods.

\subsubsection{Survey Design}
To understand which factors developers look for when selecting an \npm package, we designed a survey containing three main parts.
The first part contained questions related to the background of the participants. We asked these questions to ensure that our survey participants have sufficient experience in software development and in selecting and using \npm packages.
In this part, we asked the following questions:

\begin{enumerate}[leftmargin=.48cm]

	\item How would you best describe yourself?
	      A question with the following choices and the last choice is a free-text form: Full-time, Part-time, Free-lancer, and Other.

	\item For how long have you been developing software?
	      A selection question with the following options: $<$1 year, 1$-$3, 4$-$5, more than 5 years.

	\item How many years of JavaScript development experience do you have?
	      A selection question with the following options: $<$1 year, 1$-$3, 4$-$5, more than 5 years.

	\item How many years of experience do you have using the Node Package Manager (\npm)?
	      A selection question with the following options: $<$1 year, 1$-$3, 4$-$5, more than 5 years.

	\item How often do you search for \npm packages?
	      A question with the following options:  Never, Rarely (e.g., once a year), Sometimes (e.g., once a month), Often (e.g., once a week), Very often (e.g., everyday).

	\item Which search engine interface do you use to find relevant \npm packages?
	      A question with the following multiple choices and the last choice is a free-text form: Online search on the \npm web page (i.e., \npms), Command line search, Google or other general web search engines, and Other.

\end{enumerate}

In the second part of the survey, we had a list of statements that present seventeen factors that can affect selecting \npm packages.
In particular, we asked the question \textit{``How important are the following factors when selecting a relevant \npm package?''}
\Cref{tab:selection_factors} reports the {seventeen} factors statements.
For each statement, the table presents each factor's definition and the rationale behind asking about it.
In the survey, we asked participants to rate these statements using a Likert-scale
ranges from 1~$=$~not important to 5~$=$~very important~\cite{Oppenheim1993}.
We chose to investigate these factors for two main reasons.
First, our literature review indicated that these factors are known to impact the use and selection of \npm packages.
Second, we focued on studying factors that developers can easily observe through examining the package source code or its software repository, e.g., from the GitHub website.

\begin{table}[]
	\centering
	\caption{List of factors used in selecting a packages from the \npm ecosystem.}
	\label{tab:selection_factors}
	\scalebox{0.86}{
	\begin{tabular}{p{0.8in}|p{1.9in}|p{4.3in}}
	\toprule
	\multirow{2}{*}{\textbf{Factor}} & \multirow{2}{*}{\textbf{The survey statements}}                                  & \multirow{2}{*}{\textbf{Rationale}}                                                                                                                                                                                                                                                                                                                            \\
	                                 &                                                                                  &                                                                                                                                                                                                                                                                                                                                                                \\ \midrule
	Forks & The number of forks for the package's source code on GitHub.& The number of forks that packages receive indicates that the packages are active, and many developers are contributing to these packages~\cite{Gousios2014ICSE}.                                                                                                                                                                                    \\ \midrule

	Watchers & The number of watchers of the package's GitHub repository.& Developers can watch package repositories on GitHub so they can receive notifications about package development activities~\cite{SheoranMSR2014}. The higher the number of watchers on a package indicates that the package is well-known and used by many developers. We consider that packages with an increased number of watchers refer to \used packages. \\ \midrule

	Contributors & The number of contributors to a package's GitHub repository. & The higher the number of contributors to a package repository shows that the package is more likely to attract developers~\cite{Yamashita2016JIP}.                                                                                                                                                                                                                        \\ \midrule

	Downloads & The number of downloads the package has. & The packages that have a higher number of downloads indicate that the packages are highly selected and used~\cite{Abdellatif_IST2020}.                                                                                                                                                                                                                                  \\ \midrule

	Stars & The number of stars of a packages on GitHub. & A high number of stars that a \npm package receives on GitHub could indicate to developers that the package is more likely popular, which may attract them to use the package~\cite{Borges_JSS2018,DabbishCSCW2012}.                                                                                                                                            \\ \midrule

	Dependencies                     & The number of dependencies the npm package has.                                  & A larger number of dependencies could not attract more developers to use the packages since prior work shows that packages with a more considerable dependency may lead to dependency hell~\cite{Abdalkareem_FSE2017}.                                                                                                                                         \\ \midrule

	License                          & Wether the \npm package has a permissive or restrictive software license.        & When evaluating a package, it is also essential to consider non-functional requirements, such as the license. Using a package with no license or with a license that does not match the developer organization’s usage and policies can quickly become a problem ~\cite{MelocaMSR2018,Topopens69online}.                                                       \\ \midrule

	Documentation                    & Whether the \npm package repository has online documentation, e.,g. README file. & The package that is well-documented and has a very organized README file is more likely to be used by many developers ~\cite{Begel2013IEEESoftware,Hata2015CHASE}.                                                                                                                                                                                             \\ \midrule

	Test Code                        & Whether the \npm package has test cases written.                                 & Packages that have test code written are more likely to attract developers to use them since it indicates that the packages are well-tested~\cite{Abdalkareem_FSE2017}.                                                                                                                                                                                        \\ \midrule

	Build Status                     & The build status of the \npm package for example from Travis CI.                 & The presence of a high number of failed builds in the package repository may lead developers not to use the package~\cite{Abdellatif_IST2020}.                                                                                                                                                                                                                 \\ \midrule

	Vulnerabilities                  & If the \npm package depends on vulnerable dependencies.                          & If a \npm package is affected by vulnerabilities, it may concern developers and deter them from selecting and using the package~\cite{Abdellatif_IST2020,Abdalkareem_EMSE2020}.                                                                                                                                                                                              \\ \midrule

	Badges                           & If the package repository has badges.                                            & The presence of badges in the package repository indicates that the package is of good quality that attracts developers to use the package~\cite{trockman2018ICSE}.                                                                                                                                                                                            \\ \midrule

	Website                          & If the package has a custom website.                                             & The presence of a website for the package indicates that the package is supported by an organization, which is usually a signal that there is more than one maintainers or major contributor (i.e., there is support by an organization)~\cite{Qiu2019CSCW}.                                                                                                   \\ \midrule

	Releases                         & The release frequency of the package.                                            & A package with several releases indicates that the package is well maintained.                                                                                                                                                                                                                      \\ \midrule

	Closed Issues                    & The number of closed issues in the package’s repository.                         & The number of closed issues indicates how well-maintained the package is and reveals how maintainers of the package respond to issues. Packages with a large percentage of closed issues attract more developers to use the package~\cite{Abdellatif_IST2020}.                                                                                                 \\ \midrule

	Commit Frequency                 & The commit frequency in the package repository.                                  & Developers mainly look for well-maintained and active packages to use. Prior work also shows that the number of commits a package receives can give a good indication of how active the package is, which results in high selection~\cite{Abdellatif_IST2020}.                                                                                                        \\ \midrule

	Usage                            & The number of projects using the package on GitHub.                              & Packages that are used by many other developers are more likely to attract more developers to use~\cite{Abdalkareem_EMSE2020}.                                                                                                                                                                                                                                 \\ \bottomrule
\end{tabular}

	}
\end{table}

In the last part of our survey, we asked the participants an open-ended question about whether they had any additional comments or other factors that they look for when they select a package.
We asked this open-ended question to give our survey participants maximum flexibility to express their opinion and experience with the selection of \npm packages, which also complies with the survey design guidelines~\cite{dillman2011mail}.

Once we had our survey questions, we shared the survey with three of our colleagues who are experts in JavaScript programming and using packages from \npm.
We performed this pilot survey to discover potential misunderstandings or unexpected questions early on and improve our survey~\cite{dillman2011mail}.

\subsubsection{Participant Recruitment}
To identify the participants in our survey, we needed to reach out to developers who are the experts in selecting and using JavaScript packages.
Thus, we resorted to the public registry of \npm~\cite{registry47online}.
The registry contains information on each package published on \npm, including information about the developers maintaining the package.
We used the \npm registry to collect a list of emails and names of JavaScript developers who selected and used a sufficient number of \npm packages.
To do so, we analyzed the \npm registry, and for each package, we extracted and counted the number of its dependencies and the contacts information of the developers who are maintaining the package.
Then, we selected the top thousand developers based on the number of their distinct package dependencies.
It is important to note that we selected developers who use a large number of packages since they likely went through the process of selecting \npm packages several times.

Once we identified this initial sample of developers, we examined all the names and email addresses of the identified developers to exclude duplicated emails and names.
Based on this step, we identified {931} unique JavaScript developers.
Next, we sent email invitations of our survey to the {931} unique developers.
However, since some of the emails were returned for several reasons (e.g., invalid emails), we successfully reached 895 developers.
In the end, we received {118} responses for our survey after having the survey available online for ten days.
This number of responses translates to a 13.18\% response rate, which is comparable to the response rate reported in other software engineering surveys~\cite{SmithCHASE2013}.

\subsubsection{Survey Participants}

\Cref{tab:survey:background} shows the positions of the participants, the development experience of the participants, the JavaScript experience of the participants, and their experiences in using \npm ecosystem.

\begin{table}[]
	\centering
	\caption{Participants’ position, and their experience in software development, JavaScript, and use of the \npm package manager platform.}
	\label{tab:survey:background}
	\resizebox{\textwidth}{!}{
	\begin{tabular}{lrl|lrl|lrl|lrl}
	\toprule
	\textbf{\begin{tabular}[|c]{@{}l@{}}Developers'\\ Position\end{tabular}} & \multicolumn{2}{c|}{\textbf{Occurrences}} & \textbf{\begin{tabular}[c]{@{}l@{}}Development\\ Experience\end{tabular}} & \multicolumn{2}{c|}{\textbf{Occurrences}} & \textbf{\begin{tabular}[c]{@{}l@{}}Experience\\ in JavaScript\end{tabular}} & \multicolumn{2}{c|}{\textbf{Occurrences}} & \textbf{\begin{tabular}[c]{@{}l@{}}Experience in\\ Using \npm\end{tabular}} & \multicolumn{2}{c}{\textbf{Occurrences}}                                                          \\ \midrule
	Full-time                          & 84                                        & \sbar{84}{118}                     & ~$<$ 1                                    & 2                                  & \sbar{2}{118}                             & ~$<$ 1                             & 0                                        & \sbar{0}{118}  & ~$<$ 1          & 0  & \sbar{0}{118}  \\
	Part-time                          & 9                                         & \sbar{9}{118}                      & 1 - 3                                     & 21                                 & \sbar{21}{118}                            & 1 - 3                              & 25                                       & \sbar{25}{118} & 1 - 3           & 59 & \sbar{59}{118} \\
	Freelancer                         & 15                                        & \sbar{15}{118}                     & 4 - 5                                     & 15                                 & \sbar{15}{118}                            & 4 - 5                              & 37                                       & \sbar{37}{118} & 4 - 5           & 20 & \sbar{20}{118} \\
	Other                              & 10                                        & \sbar{10}{118}                     & ~\textgreater~5                           & 80                                 & \sbar{80}{118}                            & ~\textgreater~5                    & 56                                       & \sbar{56}{118} & ~\textgreater~5 & 39 & \sbar{39}{118} \\ \bottomrule
\end{tabular}

	}
\end{table}

As for the participants' positions, {84} participants identified themselves as full-time developers and {9} participants as part-time developers.
Interestingly, {15} participants identified themselves as freelancers.
The remaining ten participants identified themselves as having other positions not listed in the question, including open-source developers,  IT specialists, and PhD students.

Of the {118} participants in our survey, {80} participants have more than 5 years of development experience and {15} responses have between 4 to 5 years.
Also, {21} participants claimed to have between 1 to 3 years of experience, and only two participants have less than two years of development experience.
In addition, {56} participants have more than 5 years of experience in JavaScript, {37} participants have experience in using JavaScript between 4 to 5 years, and {25} participants claimed to have between 1 to 3 years of experience.

We also asked our survey participants about their experience in using packages from the \npm ecosystem.
The majority of our survey participants indicated that they have more than one year of experience using \npm.
Specifically, {39} participants have more than 5 years of experience using \npm and {20} responses have between 4 to 5 years.
Finally, {59} participants claimed to have between 1 to 3 years of experience.

\begin{figure}[b]
	\centering
	\includegraphics[width=.54\linewidth]{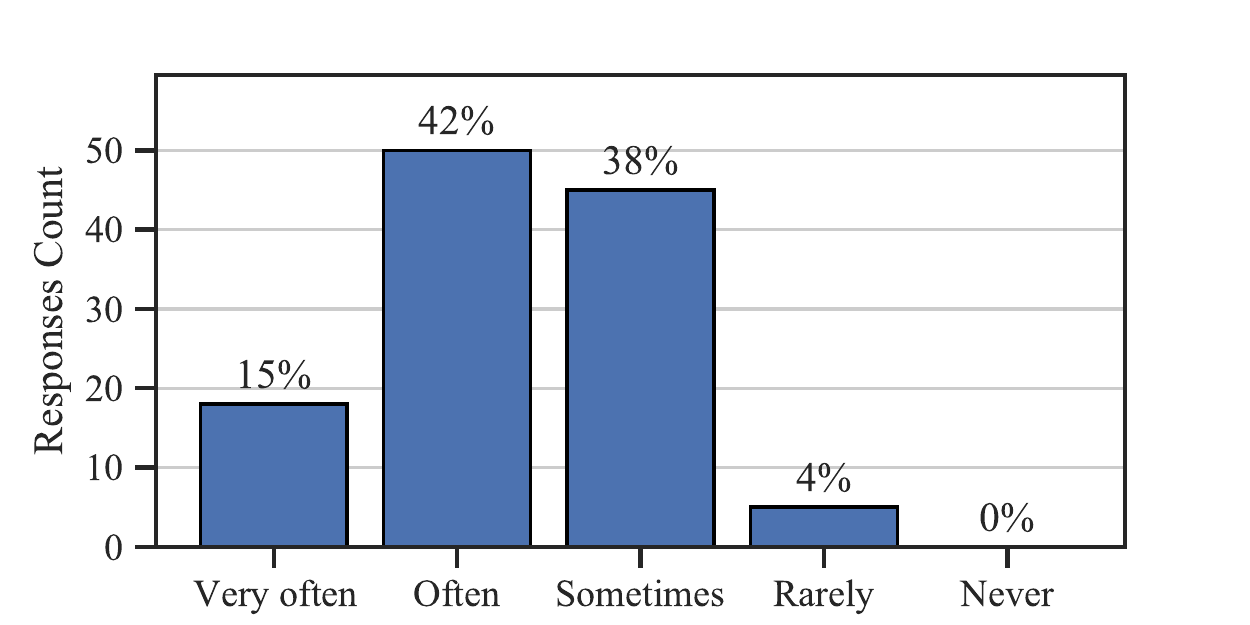}
	\fullcaption[Survey responses regarding how often our survey participants search for \npm packages.]{In our survey, the question has the following answers: never, rarely (e.g., once a year), sometimes (e.g., once a month), often (e.g., once a week), very often (e.g., everyday).}
	\label{fig:search_for_npm_packages}
\end{figure}

In addition, to inquire our survey participants about their development experience, we asked them how often they search for \npm packages and which search engine they used to perform their search.
\Cref{fig:search_for_npm_packages} reports the result related to participants’ habits about how often they search for \npm packages.
Of the 118 participants, 15\% indicated that they search for \npm packages very often, and 42\% indicated that they often search for new \npm packages.
Almost all the remaining participants (38\%) indicated they sometimes look for \npm packages.
Interestingly, only 4\% of our survey participants reported that they rarely do search for packages, and no one indicated that she/he never looks for \npm packages.
Our survey participants also reported that they mainly use web search engines (e.g., Google) when they search for \npm packages to use.
Interestingly, only 20\% of them indicated they use other search engines.

Overall, the background information about the developers who participated in our survey shows that they are experienced in JavaScript and selecting \npm packages, which gives us confidence in the finding based on their experiences.

\subsubsection{Analysis Method}

To analyze our survey responses about the different factors used to select \npm packages, we first showed the distribution of the Likert-scale for each factor, which ranges from~1~$=$~not important to~5~$=$~very important~\cite{Oppenheim1993}. Second, for all responses of each factor, we calculated values of the median, the interquartile range (IQR), the mean, and the standard deviation (SD).

In addition, to analyze the free-text answers from the open-ended question related to developers' opinions, we performed an iterative coding process to understand whether the responses indicated any other factors that we did not consider in our survey~\cite{rea2014designing}.
The first two authors iteratively developed a set of codes based on an inductive analysis approach~\cite{seaman1999qualitative}.
In total, the authors manually examined {30} responses from the developers who answered the optional open-ended question.
However, based on this analysis, we did not find any new factors that we did not consider in our survey.
In fact, all the responses to this open-ended question supported the developers' opinions about the studied factors.

\subsection{Study Results}

\Cref{tab:survey_results} shows the factors' name and the 5-point Likert-scale distribution for each factor from our survey responses.
The table also shows the scale's median alongside the value of IQR and mean alongside SD.
Overall, based on our survey results, we can divide the factors used by developers when selecting packages into three groups:
1)~important factors (e.g., documentation, downloads, and stars),
2)~somewhat important factors (e.g., license and testing), and
3)~unimportant factors (e.g., watchers and badges).
In the following, we discuss the developers' perceptions in more detail:

\begin{table}[]
	\centering
	\caption{ Survey results of the factors used in selecting a package from the \npm ecosystem.}
	\label{tab:survey_results}
	\begin{tabular}{l|c|cc|cc}
	\toprule
	\multirow{2}{*}{\textbf{Factor}} & \textbf{Distribution}                                              & \multirow{2}{*}{\textbf{Median}} & \multirow{2}{*}{\textbf{IQR}} & \multirow{2}{*}{\textbf{Mean}} & \multirow{2}{*}{\textbf{SD}} \\
	                                 & {\small1~2~3~4~5}                                                  &                                  &                               &                                &                              \\ \midrule
	Documentation                    & \includegraphics[width=.5in]{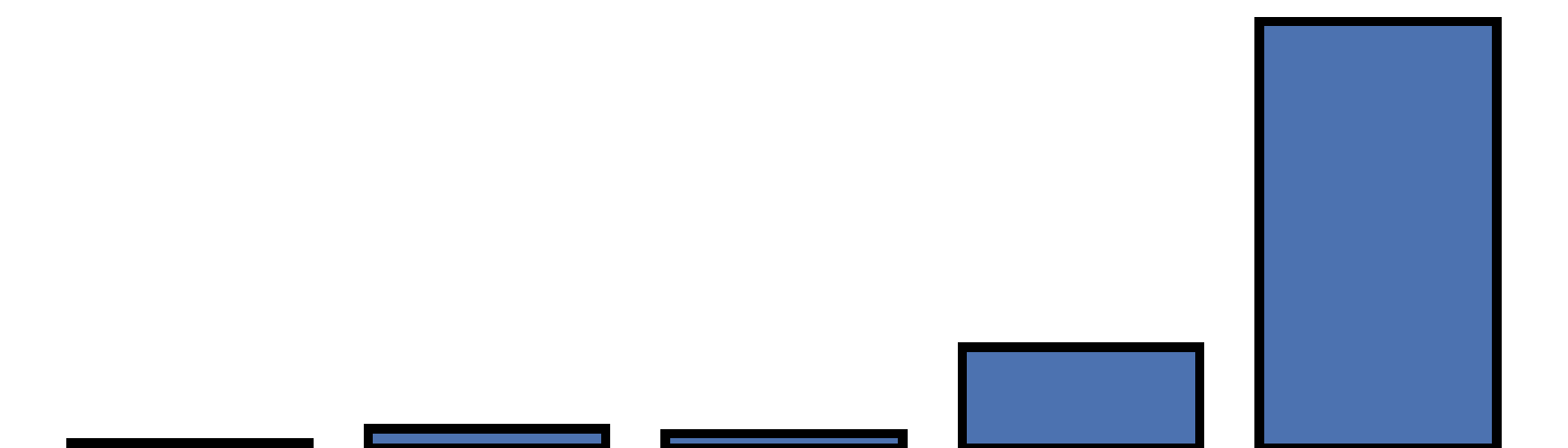}    & 5.0                              & 0.0                           & 4.64                           & 0.77                         \\
	Downloads                        & \includegraphics[width=.5in]{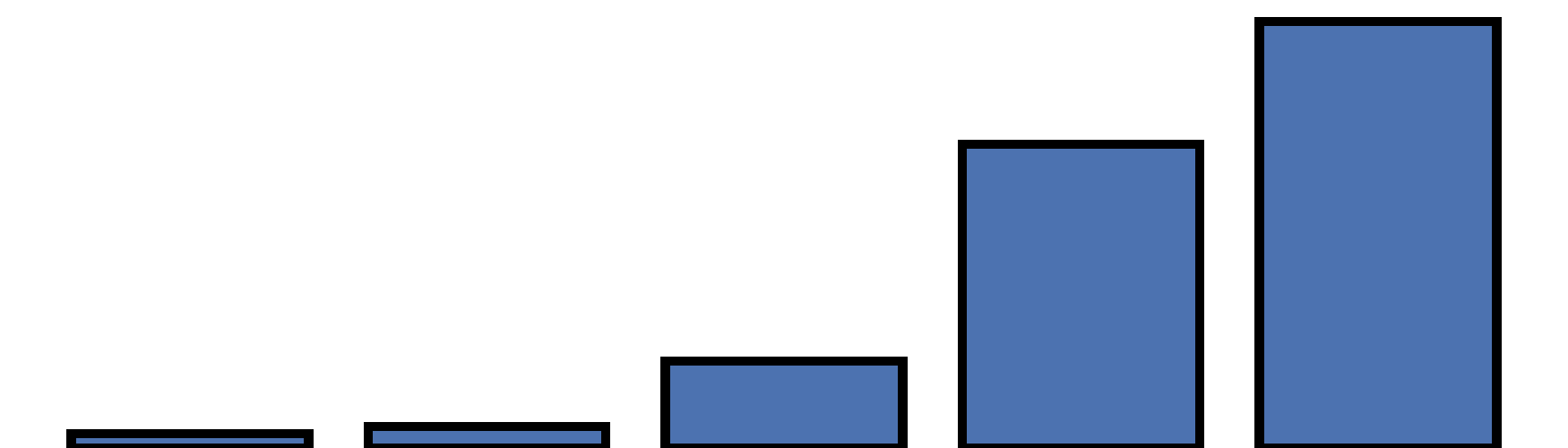}        & 4.5                              & 1.0                           & 4.30                           & 0.88                         \\
	Stars                            & \includegraphics[width=.5in]{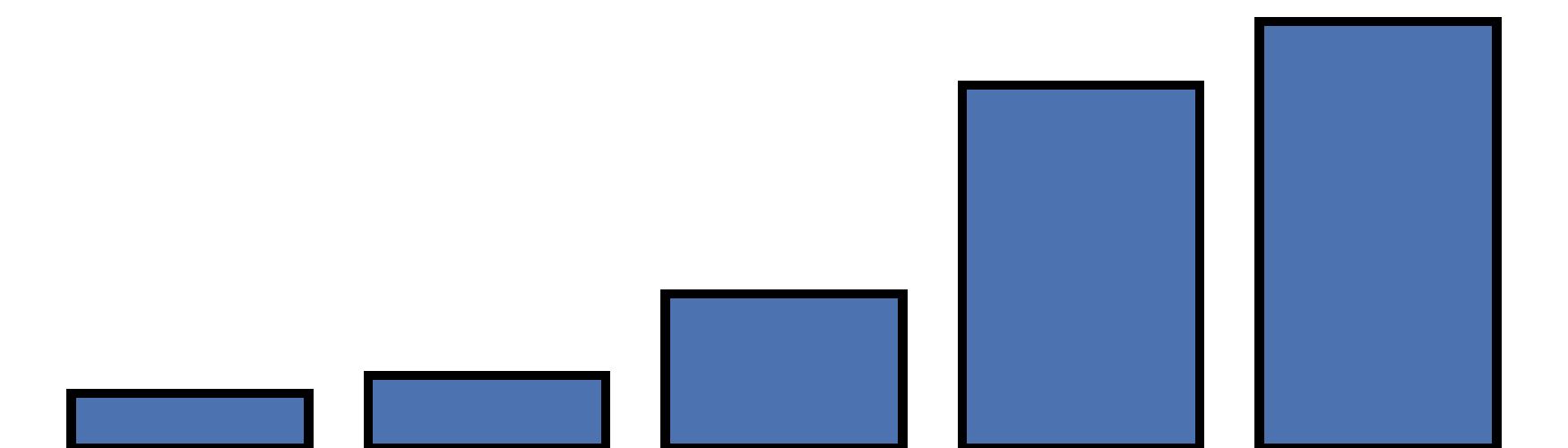}            & 4.0                              & 2.0                           & 3.97                           & 1.13                         \\
	Vulnerabilities                  & \includegraphics[width=.5in]{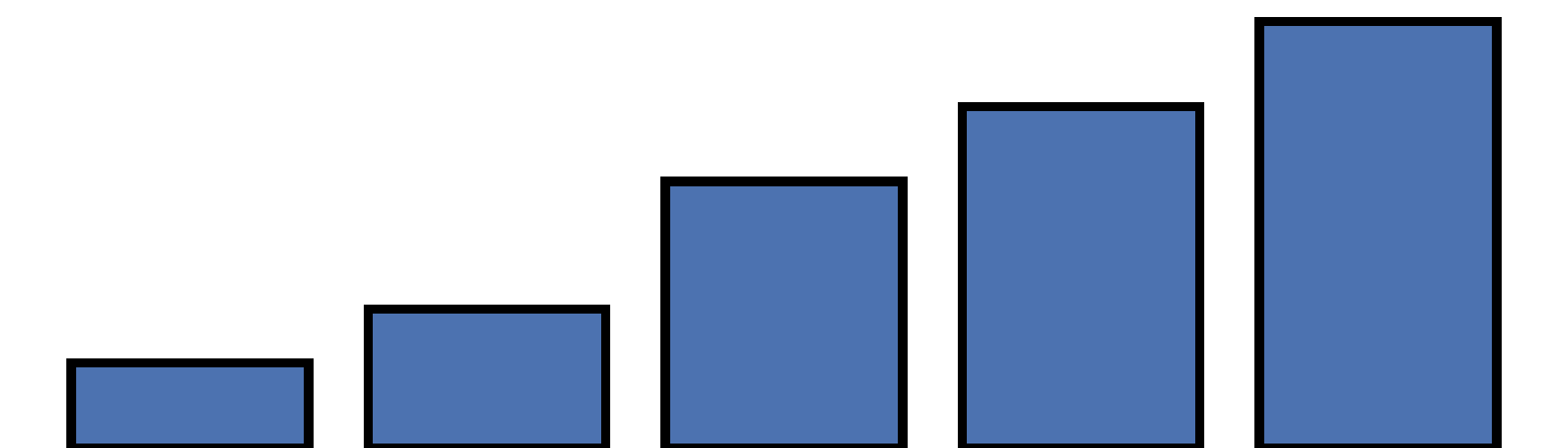}  & 4.0                              & 2.0                           & 3.70                           & 1.24                         \\
	Release                          & \includegraphics[width=.5in]{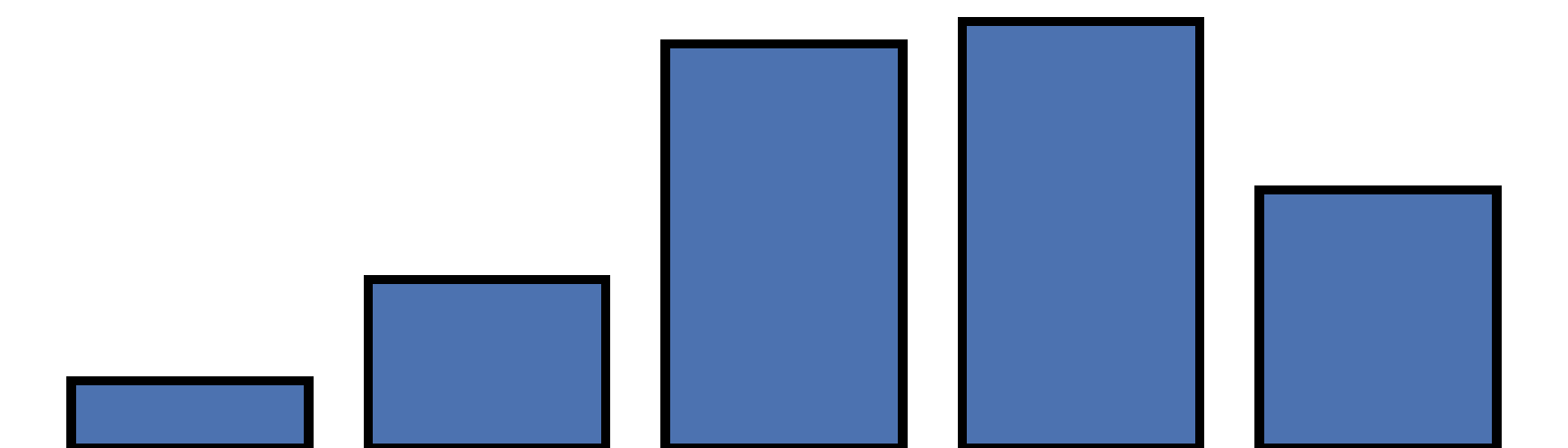}         & 4.0                              & 1.0                           & 3.48                           & 1.10                         \\
	Commit frequency                 & \includegraphics[width=.5in]{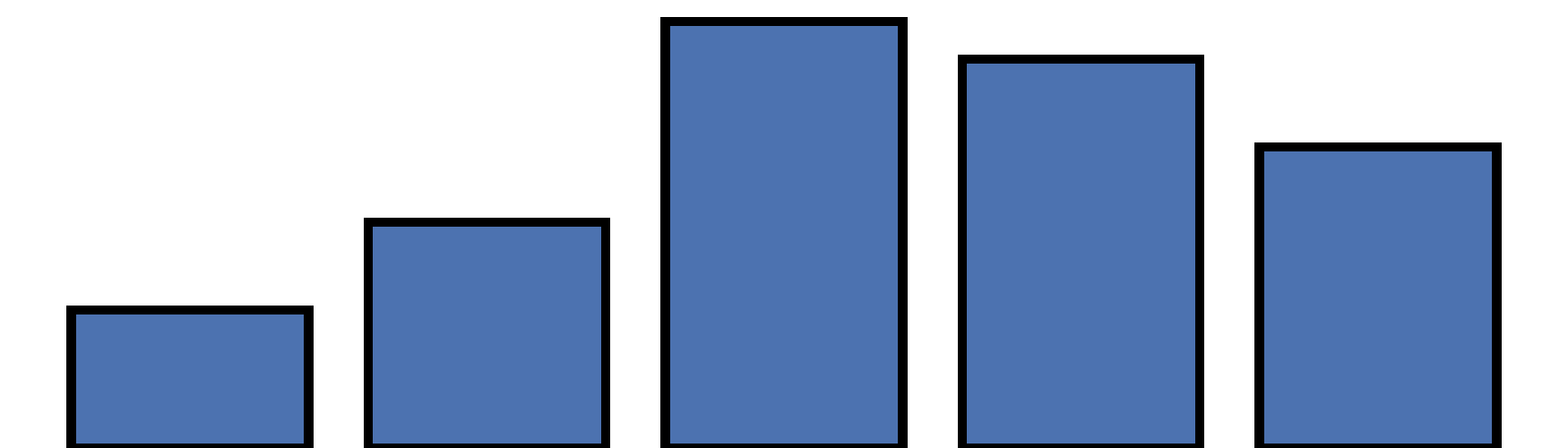} & 3.0                              & 1.0                           & 3.33                           & 1.23                         \\
	Closed issue                     & \includegraphics[width=.5in]{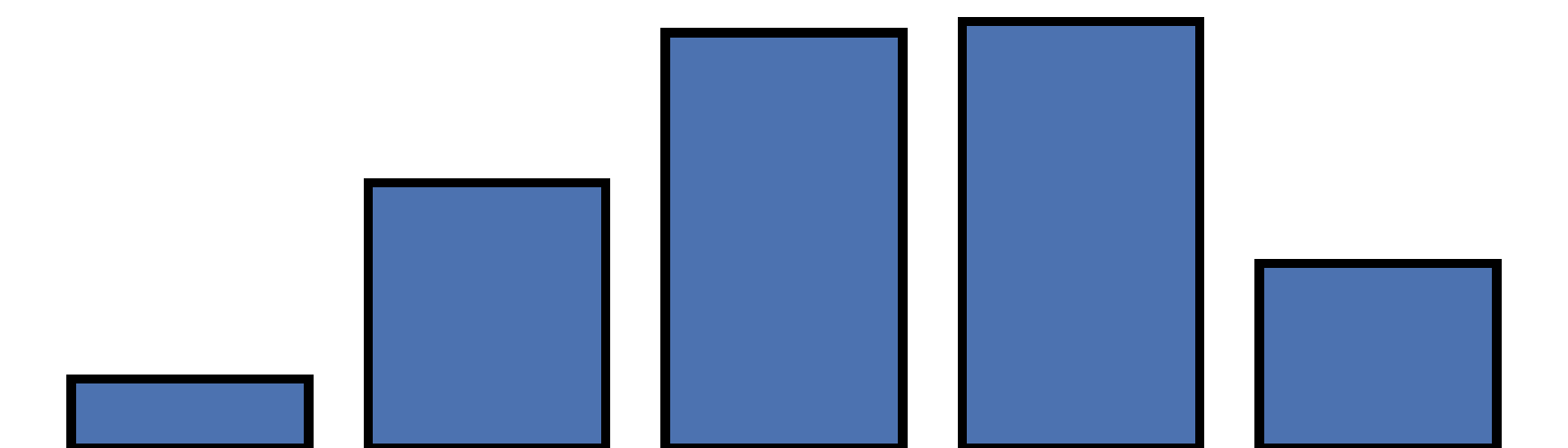}      & 3.0                              & 1.0                           & 3.29                           & 1.09                         \\
	License                          & \includegraphics[width=.5in]{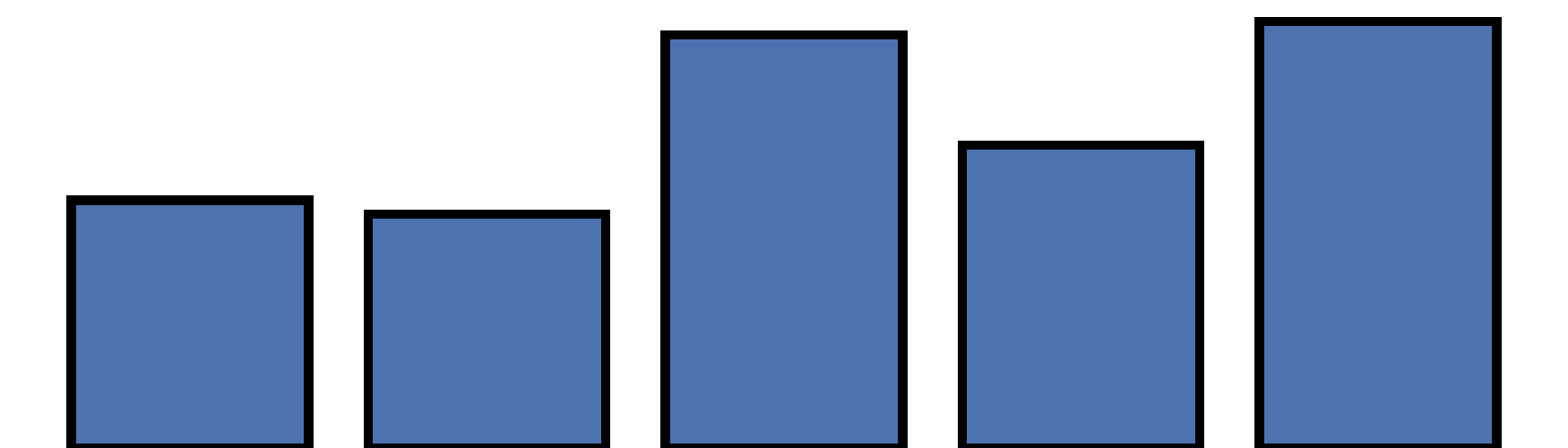}          & 3.0                              & 3.0                           & 3.26                           & 1.39                         \\
	Usage                            & \includegraphics[width=.5in]{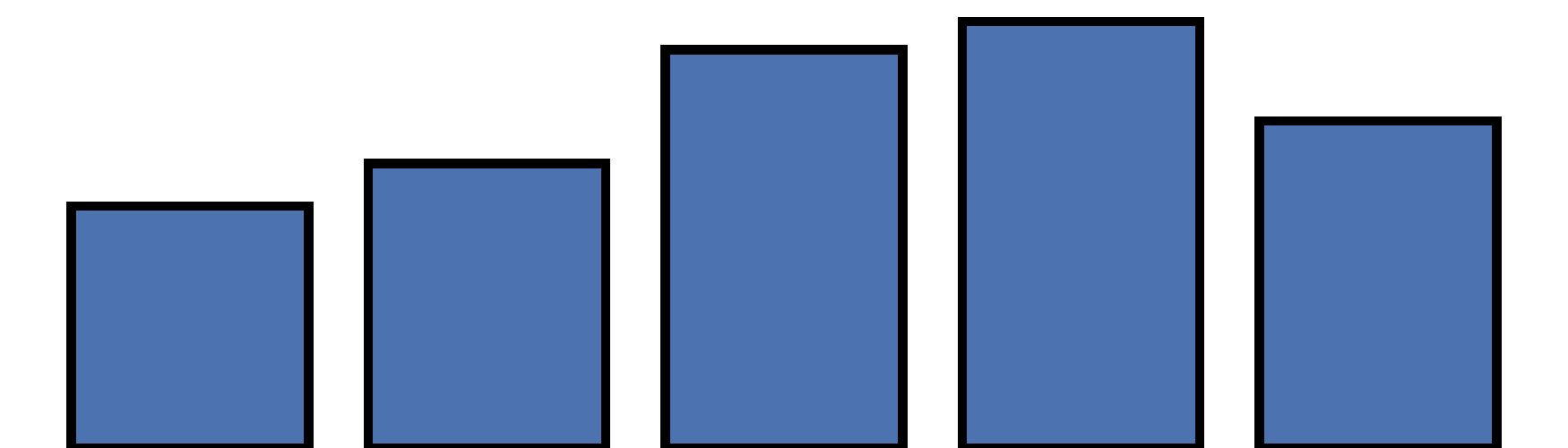}            & 3.0                              & 2.0                           & 3.19                           & 1.33                         \\
	Test Code                        & \includegraphics[width=.5in]{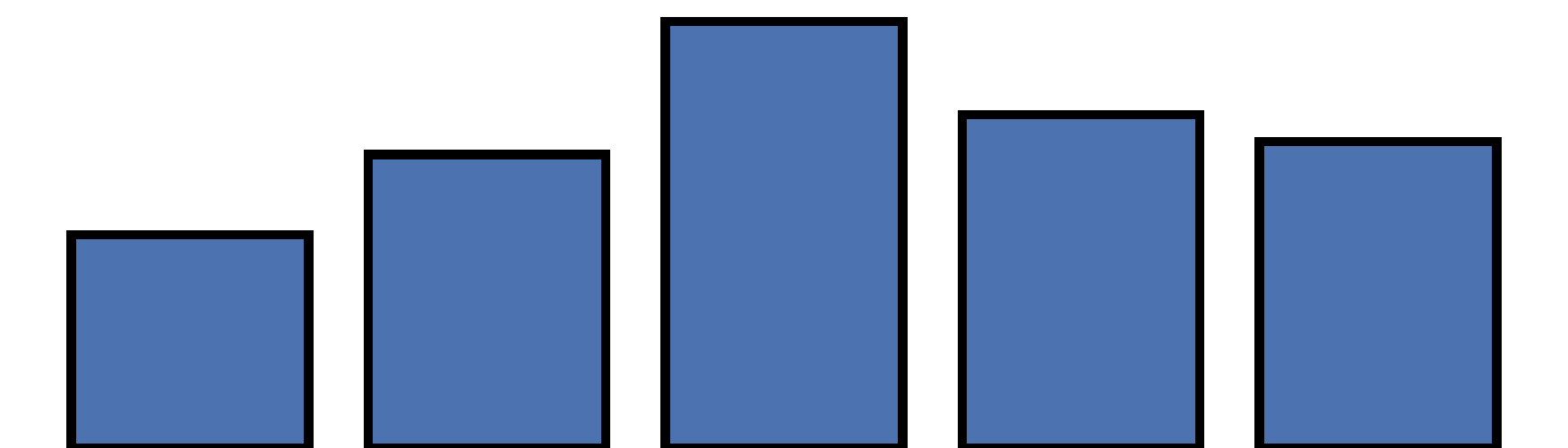}        & 3.0                              & 2.0                           & 3.14                           & 1.31                         \\
	Dependencies                     & \includegraphics[width=.5in]{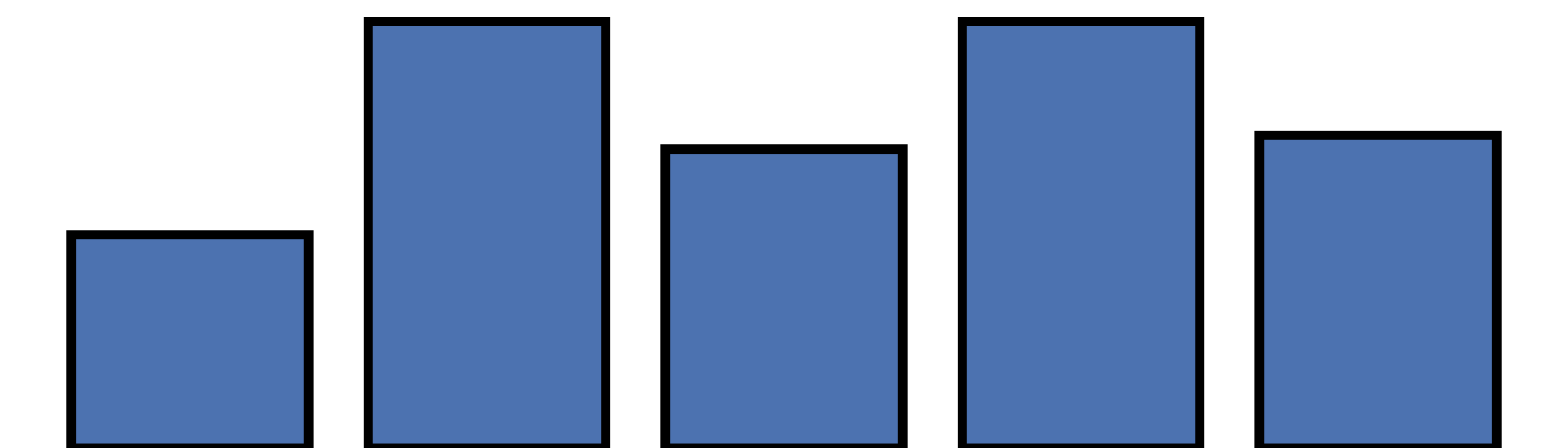}     & 3.0                              & 2.0                           & 3.12                           & 1.33                         \\
	Contributors                     & \includegraphics[width=.5in]{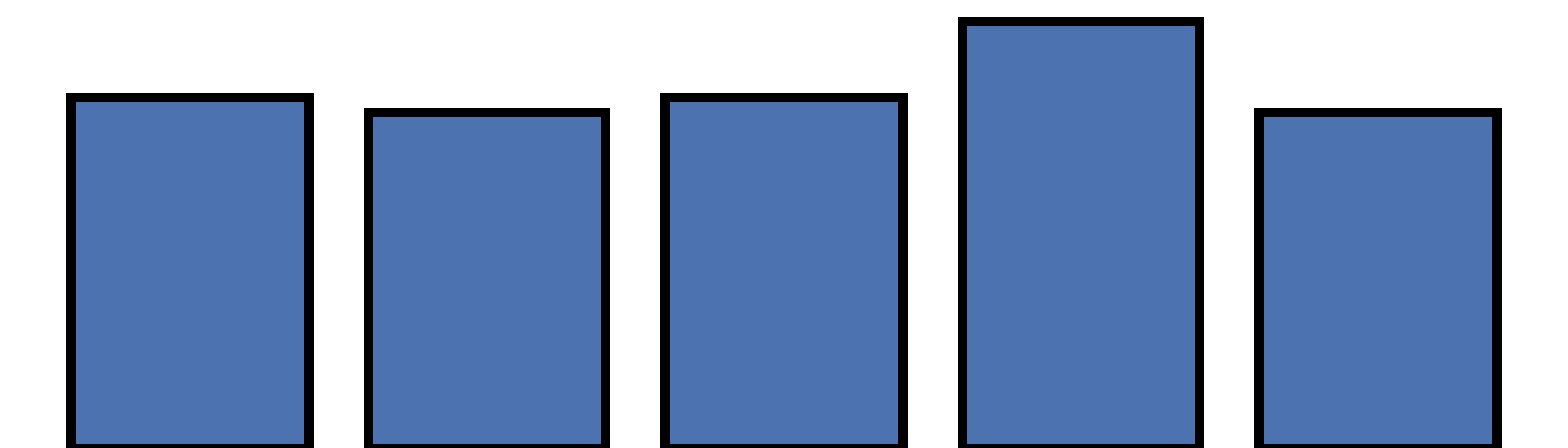}     & 3.0                              & 2.0                           & 3.03                           & 1.40                         \\
	Build Status                     & \includegraphics[width=.5in]{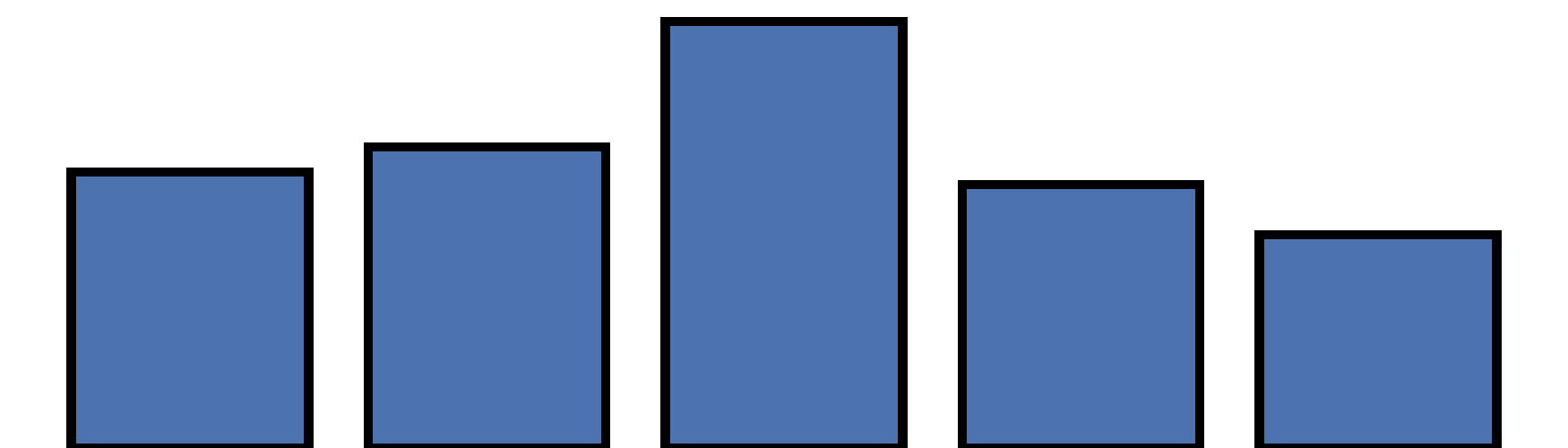}     & 3.0                              & 2.0                           & 2.89                           & 1.31                         \\
	Website                          & \includegraphics[width=.5in]{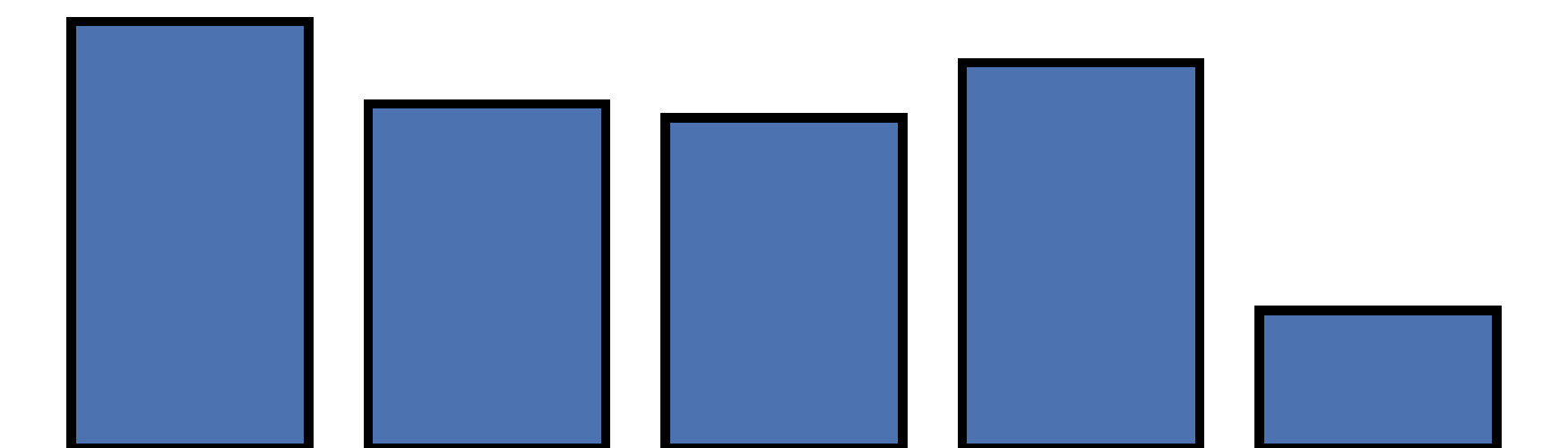}          & 3.0                              & 3.0                           & 2.67                           & 1.32                         \\
	Watchers                         & \includegraphics[width=.5in]{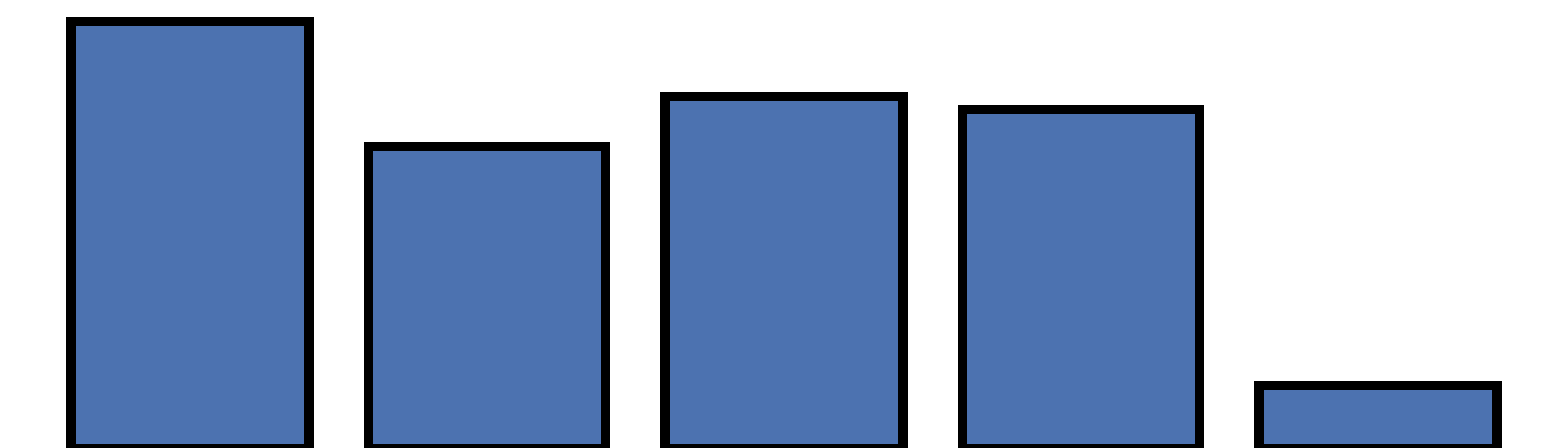}         & 3.0                              & 3.0                           & 2.53                           & 1.25                         \\
	Badges                           & \includegraphics[width=.5in]{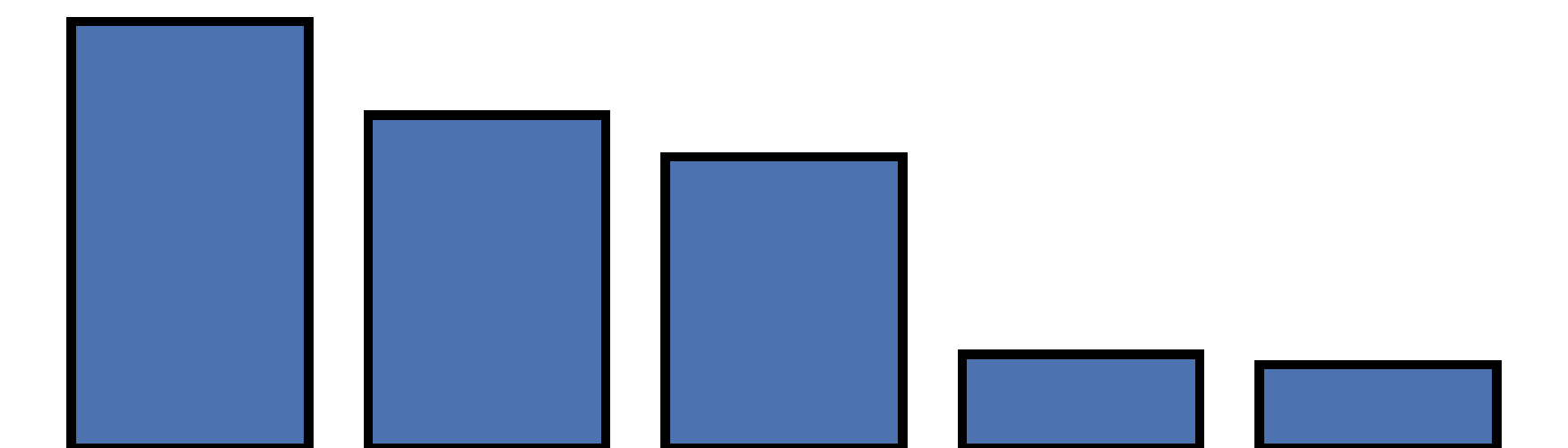}           & 2.0                              & 2.0                           & 2.25                           & 1.20                         \\
	Forks                            & \includegraphics[width=.5in]{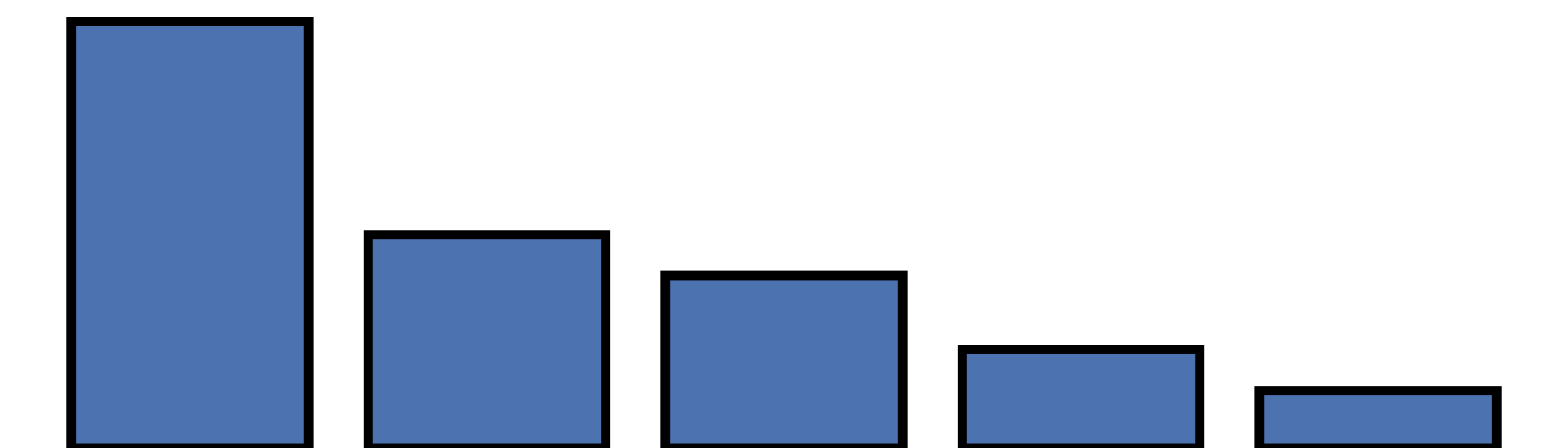}            & 2.0                              & 2.0                           & 2.12                           & 1.25                         \\
	\bottomrule
\end{tabular}

\end{table}

\textbf{Documentations:}
On a 5-point scale, participants indicated that the most important factor when looking for an \npm package to select is how well a package is documented.
\Cref{tab:survey_results} shows that the majority {93\%} (median~$=$~5.00 and mean~$=$~4.65) of the responses agree with the statement that the GitHub repository of a \npm package that they are examining to select should have some form of documentation.
In addition, to confirm this statement, developer P40 stated that \textit{``Sample code documentation on its usage''} 
are important factors when selecting an \npm package to use.

\textbf{Downloads:}
The second most important factor reported by our survey participants is the number of downloads that the packages have.
More than {85\%} (median~$=$~4.5 and mean~$=$~4.30 on the 5-point scale) of the responses said that they considered the number of downloads a package has when searching for a package to use.
These results give a high indication that developers still consider the download count of packages as a sign of the community interest, which means that the package is a good option to select.

\textbf{Stars:}
Our survey showed that developers also look for the number of stars the packages have when searching for a package to select.
On the 5-points scale, developers believed that the reputation of the packages in terms of start count is an important indicator with median~$=$~4.0 and mean~$=$~3.97.
For example, developer P74 stated that~\textit{``reputation/popularity''} are the most important factors when selecting an \npm package to use.

\textbf{Vulnerabilities:}
The fourth most important factor developers consider when searching for a new \npm package to use is that the packages do not depend on vulnerable code.
On a 5-point scale, {62\%} of the developers saw vulnerabilities as an essential factor when finding and selecting relevant \npm packages.
Furthermore, some developers in our survey emphasized the essentiality of this factor, participant P58 said~\textit{``...~not dependent on other out of date or vulnerable packages.''} 
Also, participant P45 stated that they look for packages that are free of vulnerable code and the maintainers of the packages use tools to scan for vulnerabilities such as Snyk and dependabot tools.

Our survey also revealed that there are some other factors that developers did not have an agreement on whether they are essential when they search for packages to select or not.
We found that factors such as release (median~$=$~4.0 and mean~$=$~3.48), commits frequency (median~$=$~3.0 and mean~$=$~3.33), and test code (median~$=$~3.0 and mean~$=$~3.14) do not have a consistent agreement amongst the participants in our survey.
However, some participants explicitly highlighted the importance of some of these factors, such as developer P69, who said~\textit{``examining the package repository and see the recent and historical activity/commits/updates would help making the decision''}.
Another developer, P1 explained that~\textit{
	``. . . the test coverage status is useful, but can be verified manually in the code when deciding to use the package. The last date a commit was made is very important.
	The more recent the better.
	The last date a release was made is very important.
	The more recent the better.''
}
In addition, we observed from \Cref{tab:survey_results} that developers in our survey did not have a consistent agreement about factors such as license and number of dependent applications, number of dependencies that the package uses, the number of closed issues, and the number of contributors, which have, on a 5-point scale, values with a mean of {3.26, 3.19, 3.12,  and 3.03}, respectively.

The other interesting group of the studied factors that developers tend not to consider when examining an \npm package to use are: forks, badges, watchers, website, and build status.
Our analysis showed that these factors received median values between 2.0 and 3.0 and mean values between 2.12 and 2.89 on a 5-point scale.
However, only one developer from our survey supported the idea that examining the build status is essential when selecting a package to use and said P1~\textit{``The build status is important no matter if it comes from Travis CI or other providers~...''}.

Finally, we observed that developers mentioned a few other factors when looking for \npm packages.
Our survey participants indicated that if there is a big software company that supports the package.
For example, developers P39 said~\textit{``The source of the package, if it is by a company that actively supports open source and maintains their open source packages (ex: Facebook, Formidable labs, Infinite Red), brings more points''}.
Also, another participant stated the same, P11~\textit{``Private support for big companies in open source projects or libs (angular-google, react-facebook, etc) that means the package usually follow good practices, test, linter, ci, etc, and the team that maintains the package is really good.''}

In addition, two other developers in our survey indicated that support of community discussions about the packages matters.
For example, P94 mentions~\textit{``Whether the package is actively maintained by developers well known and reputed in the community \& whether the package has good typescript support''} and P60 said~\textit{``If the library is supported with an online community where usage is discussed''}.
Another developer, P20 stated~\textit{``References on other professional webpages about the package''}.

\conclusion{
	In summary, JavaScript developers have access to a wealth of information about a large number of \npm packages that can be used when deciding which packages to select.
	Our survey shows that developers mainly consider packages that are well-documented, popular, and do not suffer from security vulnerabilities.
	Moreover, when we conducted our survey, among the 118 respondents, 73 (62\%) provided their emails and showed interest in our findings.
	This indicates the strong relevance and importance of the findings to the practitioners and the overall JavaScript development community.
}

\section{Quantitative Analysis}
\label{sec:survey:quantitative_analysis}
The goal of this analysis is to triangulate our qualitative findings.
In particular, we wanted to quantitatively validate the developers' perception about the factors that \used \npm packages possess.
In this analysis, we examined {2,592} \npm packages divided into \used and \unused packages.
For each package in our dataset, we collected quantitative data to present the factors studied in our survey.
Then, we used regression analysis to quantitatively investigate which of the studied factors are the most important.

\subsection{Study Design}
In this section, we described our methodology of collecting a dataset of \used and \unused \npm packages.
We also described how we collected the studied factors, which served as the dependent variables in our study.
Finally, we presented our analysis method and steps.

\subsubsection{Data Collection}
To quantitatively examine the factors that make some \npm packages \used, we wanted to have a sufficient number of packages that present both \used and \unused packages.
To do so, we resorted to study packages from the \npm ecosystem.
We started by retrieving the metadata information of all the \npm packages that are published on the \npm ecosystem.
In particular, we wrote a crawler to interact with the \npm registry and download the package.json file of every \npm package~\cite{registry47online}. %
It is important to note that the {package.json} contains all the package information, including the names of other packages that the package depends on them.
Once we have the {package.json}, we started recursively analyzing the package.json file of every package to extract its dependencies.
After that, for each package in the \npm ecosystem, we counted the number of other packages that are listed as dependencies, i.e., the number of dependent packages.

It is important to note that we chose to use the number of dependent packages as a proxy of \used packages from the \npm ecosystem over other measurements, particularly download count, for two main reasons.
First, the \npm provides only an accumulated download count over time, which does not show the current stats of the package. %
Second, the download count that \npm provides could include crawlers and downloads due to transitive dependencies.
Furthermore, in our process, we considered only the direct dependent packages from the \npm registry and avoided including dependent applications from open-source hosting services like GitHub. 
We did this since our goal is to proxy how many times a developer went through the process of selecting a package and decided to select the subject package.
Moreover, platforms like GitHub hosts millions of applications created using predefined project templates or bootstrapping tools.
For example, the tool \textit{Create React App} alone bootstrapped millions public applications on GitHub, which may not be completed.
Considering such applications from GitHub will amplify the decision taken by the creators of such tools or templates to overtake the decisions of millions of developers.
In addition, we did not consider the number of transitive dependents because it does not reflect how many times developers have selected a package.

\begin{figure}[t]
	\centering
	\includegraphics[width=.7\linewidth]{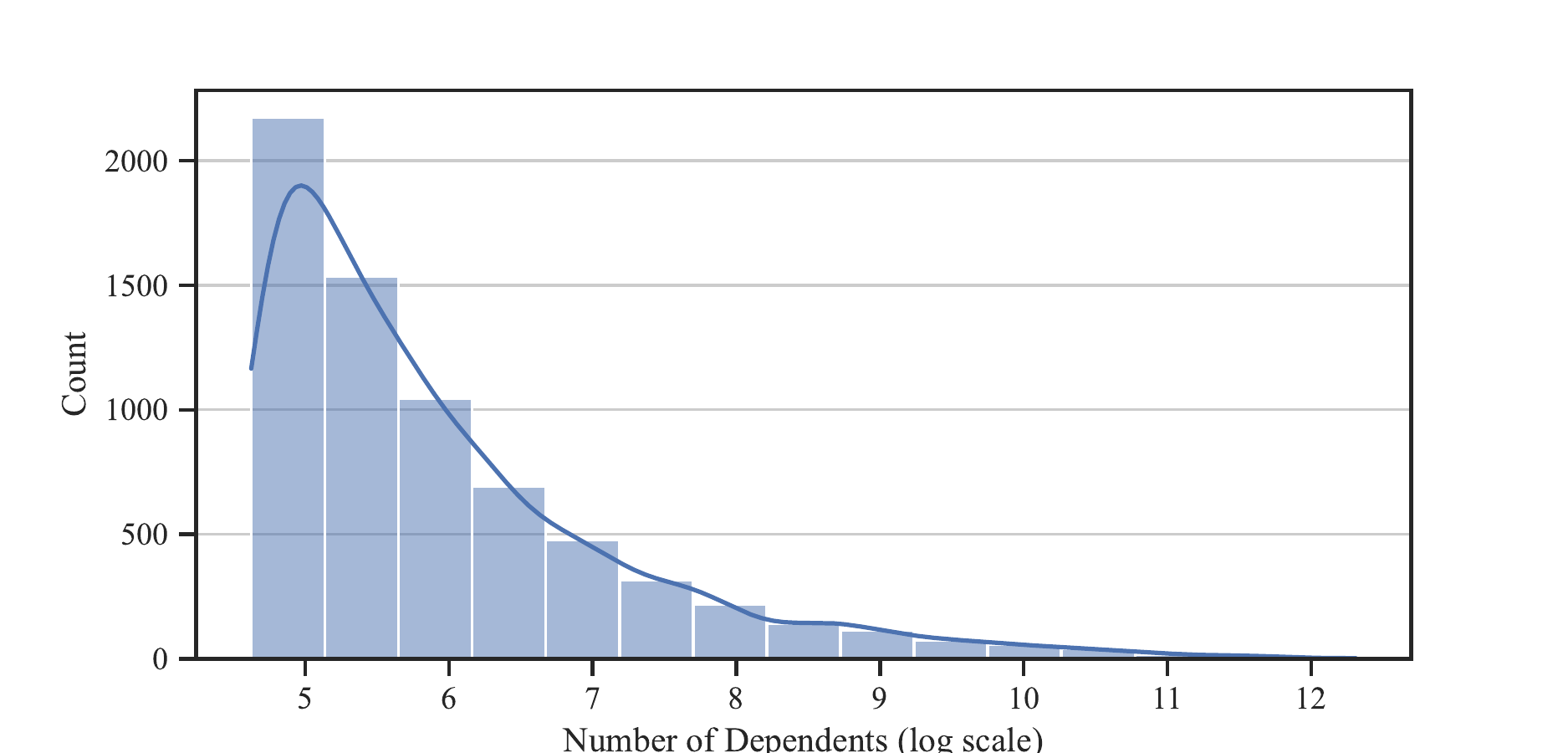}
	\caption{A histogram for the used \npm packages.}
	\label{fig:studied_packages}
\end{figure}

In total, we analyzed the package.json file of 1,423,956 \npm packages.
After that, we chose to study {6,924} packages that have more than 100 dependent packages, i.e., the number of packages that depend on the selected packages.
We decided to study \npm packages that have more than 100 dependent packages for two main reasons.
First, we found that prior work indicated that \npm ecosystem has many packages that are not used, e.g., toy packages~(e.g.,~\cite{Zerouali_SANER2019,Abdalkareem_FSE2017}).
Thus, selecting packages with more than 100 direct dependent packages eliminated incompetent packages.
Second, since we wanted to examine \used \npm packages, we focused on packages that can potentially be used and appear as an option for developers when searching for an \npm package to use, for example, packages that are widely adopted by other packages.
Moreover, we selected this threshold after examining the distribution of the number of dependent packages across all packages in the \npm ecosystem.

Next, we sorted the selected \npm packages based on their number of dependent packages.
\Cref{fig:studied_packages} presents the distribution of the number of dependent packages.
We considered the top 20\% based on the number of dependent packages as \used packages and the bottom 20\% as \unused packages.
We resorted to using these thresholds to have an essential distinction between the two samples, which element gray area between them.
Also, prior studies used a similar sampling technique~\cite{BavotaTSE2015,lee2020empirical,Tian2015ICSME}.
In the end, we had {1,385} \used packages and {1,385} \unused packages.
We used these packages in our quantitative analysis.

\subsubsection{Package Usage Factors}

Since we wanted to use regression analysis to understand the most important factors in determining \used packages, we collected {seventeen} package factors. These factors are based on the ones we studied in the qualitative analysis.
Since these factors present information that developers can observe by examining online sources about the \npm packages, we resorted to extracting these factors from four different sources:
1) GitHub, which presents the package's source code and other development activities such as issues and commits,
2) \npm, which contains information about \npm packages that developers can examine on the \npm website,
3) \npms, which is the official search engine used by the \npm platform and provides metadata about the packages, and
4) Snyk, which is a service that provides a dataset of vulnerable \npm packages and their versions.
\Cref{tab:selection_factors_for_regression_analysis} shows the factors with their names, value types, and descriptions.
In the following, we presented the detailed process of extracting the studied factors from each data source:

\begin{table}[]
	\centering
	\caption{List of factors values with their description.}
	\label{tab:selection_factors_for_regression_analysis}
		\begin{tabular}{l|l|l}
	\toprule
	\textbf{Factor}  & \textbf{Type} & \textbf{Description}                      \\ \midrule
	Forks            & Number        & Forks count on GitHub                     \\
	Watchers         & Number        & Watcher count on GitHub                   \\
	Contributors     & Number        & Contributors count on GitHub              \\
	Downloads        & Number        & Downloads count from \npm                 \\
	Stars            & Number        & Stars count on GitHub                     \\
	Dependencies     & Number        & Count of dependencies from package.json   \\
	License          & Boolean       & Weather has a permissive license          \\
	Documentation    & Number        & Size of README file                       \\
	Test Code        & Boolean       & Whether has a test script                 \\
	Build Status     & Number        & Percentage of failed jobs on last commit  \\
	Vulnerabilities  & Number        & Percentage of vulnerable versions         \\
	Badges           & Number        & Count of badges in the README file        \\
	Website          & Boolean       & Weather has a website                     \\
	Releases         & Number        & Frequency of releases                     \\
	Closed Issues    & Number        & Count of closed issues on GitHub          \\
	Commit Frequency & Number        & Count of commits  in the last year        \\
	Usage            & Number        & Count of dependent repositories on GitHub \\ \bottomrule
\end{tabular}

\end{table}

\textbf{GitHub:}
to collect the repository level factors, we used the official GraphQL API~\cite{GitHubGr34online} to collect the number of {forks}, {watchers}, {stars}, and {closed issues} for each \npm package in our dataset.
Since GraphQL API does not provide direct access to the number of contributors and build status of each package repository, we used the GitHub REST API~\cite{GitHubRE56online} to count the number of {contributors} and the list of {build status}.
In addition, to measure the commits frequency, we cloned the GitHub repositories for each of the studied packages and counted the number of commits on all branches that were committed in the latest year.
Finally, we wrote a web crawler to collect the package usage factor from the GitHub web interface, which presents the number of other GitHub repositories that depend on the package.

\textbf{\npm:}
from the official \npm registry, we retrieved the list of releases for each package in our dataset.
Then, we calculated the release frequency factor by dividing the number of releases by the number of days.
Likewise, to present the dependencies factor, we used the registry to count the number of dependencies that a package uses in its last version.
Also, to calculate the documentations factor for each package, we considered the size of the readme file.
We then measured its size in terms of the number of its characters.

Also, we used both the \npm registry and GitHub GraphQL API consecutively to retrieve the name of the license that a package declares.
We then classified the licenses into three categories:
1)~permissive licenses,
2)~semi-permissive licenses, and
3)~restrictive licenses~\cite{Topopens69online}.
Finally, we retrieved the list of {badges} for each package using a tool called {detect-readme-badges}\footnote{https://www.npmjs.com/package/detect-readme-badges}. Once we had the list of badges, we calculated the badges factor by counting the number of badges used by the package.

\textbf{\textit{npms}:}
for the download factor, we used the official \npm search (\textit{npms}\footnote{https://npms.io}) through its API to collect the number of {downloads}.
Next, we examined whether the package has test code to represent the {test code} factor.
Additionally, we used the \npms API to determine the {website} factor.
To do so, we extracted the website URL for each package in our dataset.
Since some packages refer to their GitHub repository as their main website, we filter out those URL addresses.

\textbf{Snyk:}
to collect the {vulnerabilities} factor for each package in our dataset, we wrote a web crawler to collect the list of vulnerable releases from the Snyk web interface.
Then, we divided the number of vulnerable releases by the total number of releases for each package to calculate the vulnerabilities factor.

\Cref{tab:selection_factors_for_regression_analysis} shows the name, value type, and description of the factors that we used to build our logistic regression models.
Since some packages do not have values for some factors, we filter out these packages from our dataset.
In the end, we were able to collect factor values for 1,332 \used packages and 1,195 \unused packages.

\subsection{Analysis Method}
\label{sub:quantitative_analysis_method}

To quantitatively examine the most impactful factors that determine \used packages, we used logistic regression analysis.
In our study, we examined the studied {2,527} packages, which we classified into \used and \unused packages.
We then built a logistic regression to model the dependent variable, whether a package is \used or \unused.
In the following sections, we described the steps used to build the logistic regression model.

\subsubsection{Correlation Analysis}
\begin{figure}[]
	\centering
	\includegraphics[width=.7\linewidth]{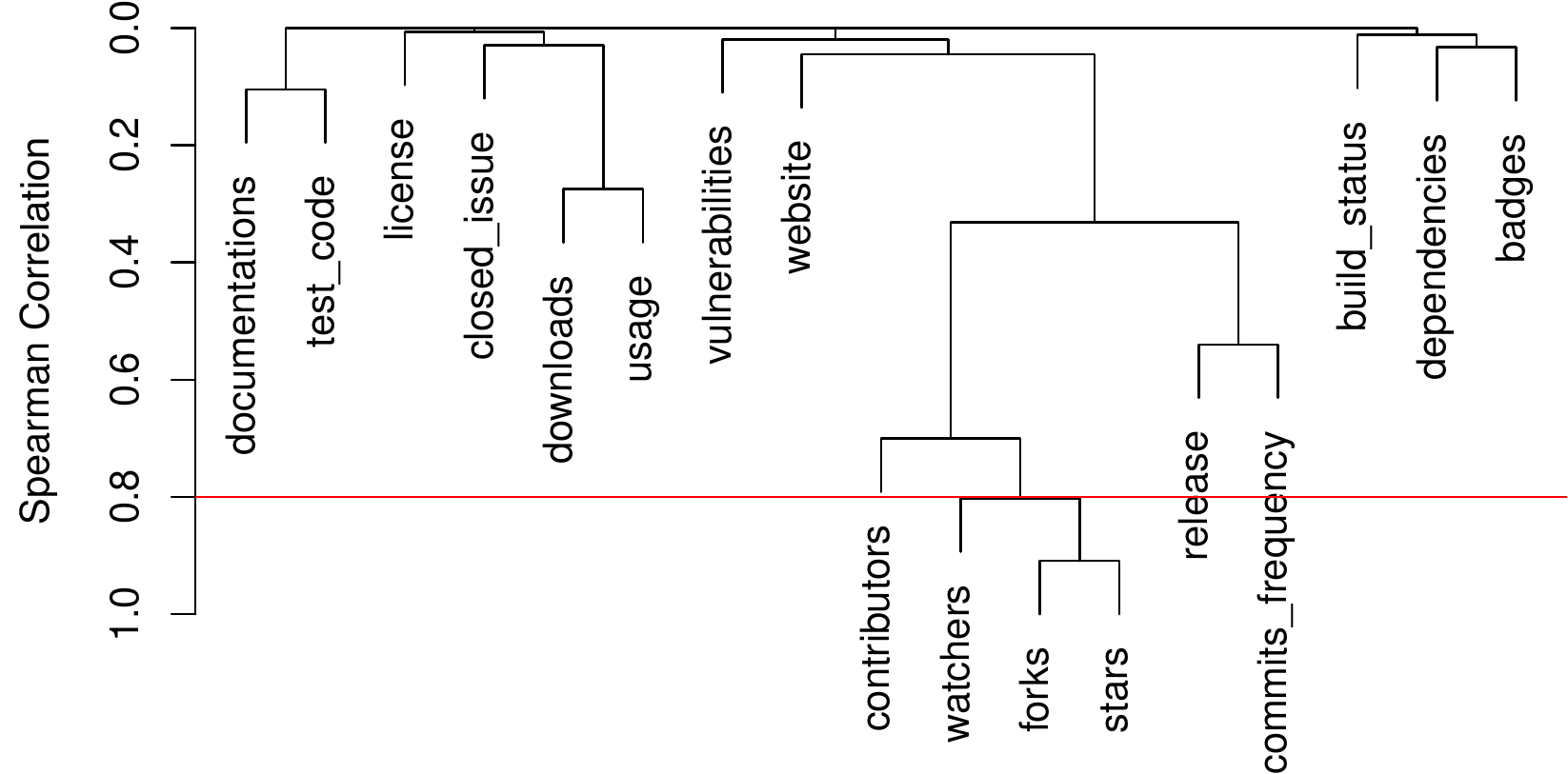}
	\fullcaption[The hierarchical clustering shows the factors that might impact the selection of \npm packages.]{
		We apply the Spearman correlation test and use a cut-off value of 0.8, to eliminate highly correlated factors.
		This analysis left us with {fifteen} factors that is used in the regression analysis.
	}
	\label{fig:cut_off_correlation}
\end{figure}

Since the interpretation of the logistic regression model can be affected by the highly correlated factors~\cite{midi2010collinearity}, we first started by removing highly correlated factors in our dataset.
Thus, we computed the correlation among the independent variables using Spearman's rank correlation coefficient.
We used Spearman correlation because it is resilient to non-normally distributed data, which is the case for our independent variables~\cite{kendall1938new}.
We considered any pair of independent variables that have a Spearman's coefficient of more than 0.8 to be highly correlated.
We selected the cutoff of 0.8  Spearman since prior work suggested and used the same threshold for software engineering data (e.g.,~\cite{Tian2015ICSME,li2017towards}).
\Cref{fig:cut_off_correlation} shows the hierarchical clustering based on the Spearman correlation among our independent variables.
From \Cref{fig:cut_off_correlation}, we observed that three factors are highly correlated, which are stars, forks, and watchers.
Finally, for these three factors, we only kept the factor that is easy to interpret, which is the number of stars.
After this analysis, we ended up having fifteen unique variables.

\subsubsection{Redundancy Analysis}
Once we removed the highly correlated factors, we also applied redundancy analysis to detect variables that do not add information to the regression analysis~\cite{harrell2015regression}.
Thus, we removed them so they do not affect the interpretation of our logistic regression model.
In our dataset, we did not find redundant variables among the remaining fifteen factors.

\subsubsection{Logistic Regression}
\label{sec:logistic_regression}
To build our logistic regression model, we followed steps that have been applied in prior studies (e.g.,~\cite{lee2020empirical}).
After identifying the factors that may impact the selection of an \npm package, we used logistic regression to model the \used packages.
Since prior studies showed that using logistic regression may be affected by the estimated regression coefficient~\cite{harrell2015regression,lee2020empirical}, we trained our model using several bootstrap iterations.
Similar to prior work (e.g.,~\cite{lee2020empirical}), we created 100 rounds of bootstrap samples with a replacement for training and testing sets that ensure the testing samples were not included in the training set and vice versa.
Then, we built a logistic regression model on the created bootstrap training samples, one for each iteration (i.e., 100 times) and test it on the testing samples.
In the end, we calculated the mean of the sample statistics out of the 100 bootstrap samples.%

\subsubsection{Evaluating Performance}
Once we built our logistic regression model, we next wanted to examine the performance of the built model.
Hence, we used the area under the receiver operating characteristic curve (ROC-AUC), which is an evaluation measurement known for its statistical consistency~\cite{bradley1997use}.
An ROC-AUC value ranges between 0 and 1, where 1 indicates perfect prediction results, and 0 indicates completely wrong predictions.
Accordingly, prior studies showed that achieving a 0.5 ROC-AUC value indicates that the model's predictions are as good as random.
However, a ROC-AUC value equal to or more than 0.7 indicates an acceptable model performance for software engineering datasets~\cite{NamASE2015,LessmannTSE2008,YanTSE2019}.
In our study, the logistic regression model achieved an ROC-AUC value of 0.74.

\subsubsection{Factors Importance}
To investigate which of the examined factors are the most impactful in our logistic regression modeling of \used packages, we used the $\textrm{Wald }\chi^{2}$ maximum likelihood tests value of the independent factors in our model~\cite{harrell2015regression}.
The higher the $\textrm{Wald }\chi^{2}$ statistics value of an independent factor, the greater the probability that its impact is significant.

We also generated nomogram charts to present the studied factors' importance on our logistic regression model~\cite{harrell2015regression,iasonos2008build}.
Nomograms are easily explainable charts that provide a way to explore the explanatory power of the studied factors.
Since $\textrm{Wald }\chi^{2}$ test provides us with only the explanatory power, we used the nomogram to show us the exact interpretation of how the variation in each factor influences the outcome of the regression model.
Furthermore, the $\textrm{Wald }\chi^{2}$ does not indicate whether the studied factors have positive or negative roles in determining \used packages or not, while the nomogram provides such information.

\Cref{fig:nomgram} shows the generated nomogram of the logistic regression model.
The line against each factor in the figure presents the range of values for that factor.
We used the points line at the top of the figure to measure the volume of each factor contribution, while the total points line at the bottom of the figure presents the total points generated by all the factors.
In our analysis, the higher the number of points assigned to a factor on the x-axis (e.g., the number of stars has 100 points), the larger its impact is on the logistic regression model.

\subsection{Study Results}
\label{sub:quantitative_result}

\begin{figure}[t]
\centering
\includegraphics[width=0.95\linewidth]{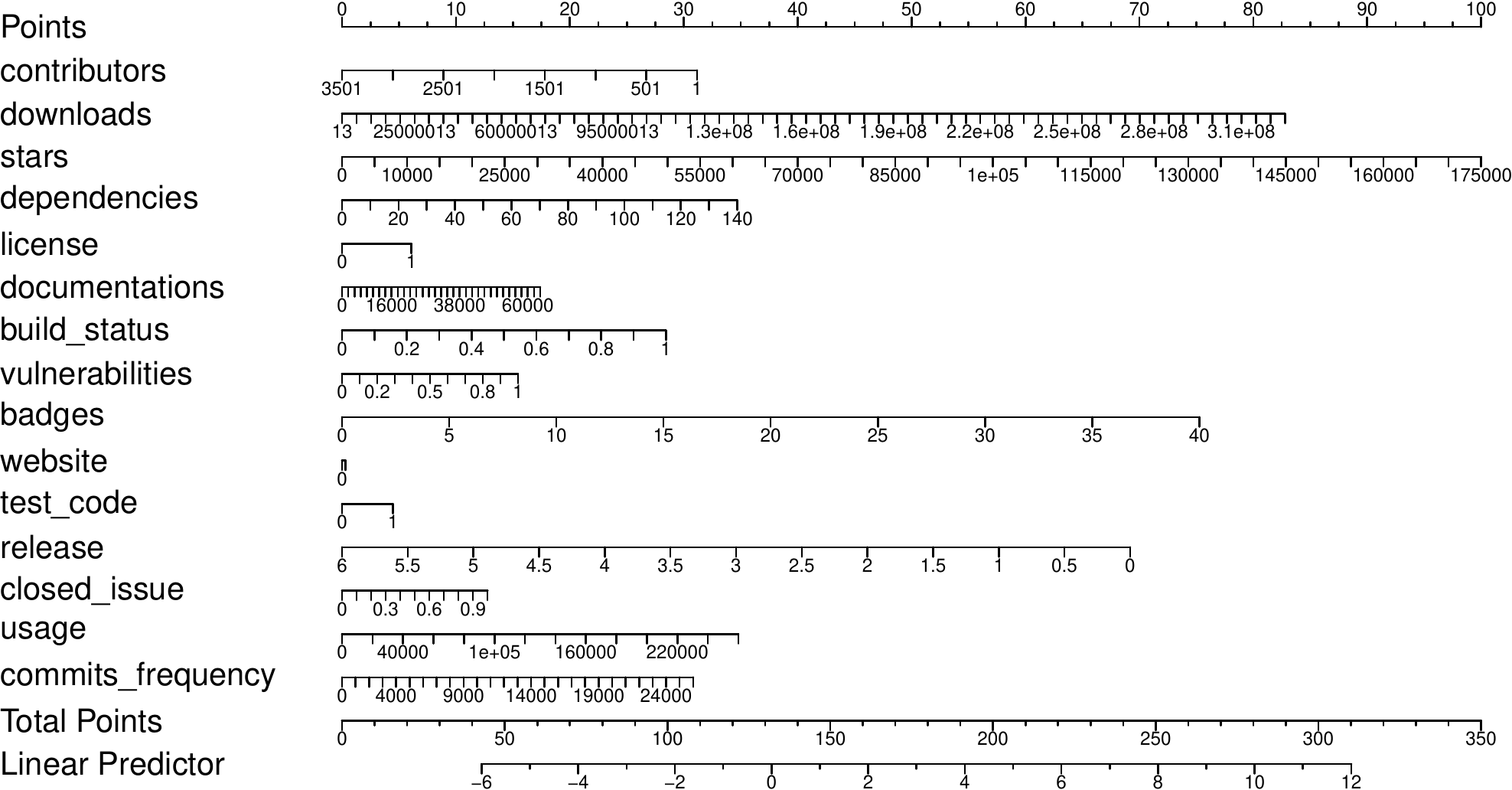}
\caption{The nomogram visually presents the impact of each of the studied factors in determining \used \npm packages.The logistic regression model used to generate this nomogram achieved a median ROC-AUC  of 0.74 on 100 out-of-sample bootstrap iterations.}
	\label{fig:nomgram}
\end{figure}

\Cref{tab:modeling} shows the values of the $\textrm{Wald }\chi^{2}$ and the $p\textrm{-value}$ for the selected fifteen factors that may impact the \used \npm packages.
\Cref{fig:nomgram} also shows the estimated effect of our factors using nomogram analysis~\cite{iasonos2008build}.
Overall, we observed that the regression analysis complemented the main qualitative findings.
However, it controverted with the importance of some factors.

\begin{table}[t]
	\centering
	\caption{The result of our logistic regression analysis for investigating the most important factors.}
	\label{tab:modeling}
	\begin{tabular} {l|rrl}
	\toprule
	\textbf{Factors}  & \textbf{$\textrm{Wald }\chi^{2}$} & \multicolumn{2}{l}{\textbf{$p\textrm{-value}$}}       \\ \midrule
	Downloads         & 63.00                             & 0.000                                           & *** \\
	Stars             & 24.21                             & 0.000                                           & *** \\
	Closed Issue      & 17.62                             & 0.000                                           & *** \\
	Vulnerabilities   & 16.47                             & 0.000                                           & *** \\
	Badges            & 12.21                             & 0.001                                           & *** \\
	Documentation     & 11.61                             & 0.001                                           & *** \\
	Dependencies      & 8.04                              & 0.005                                           & **  \\
	Build Status      & 5.54                              & 0.019                                           & *   \\
	Test Code         & 4.62                              & 0.032                                           & *   \\
	Contributors      & 4.41                              & 0.036                                           & *   \\
	Commits Frequency & 2.34                              & 0.126                                           &     \\
	Release           & 2.32                              & 0.127                                           &     \\
	License           & 1.68                              & 0.196                                           &     \\
	Usage             & 1.15                              & 0.283                                           &     \\
	Website           & 0.02                              & 0.880                                           &     \\ \bottomrule
	\multicolumn{4}{l}{\multirow{2}{*}{\footnotesize{Signif. codes:  0 `***' 0.001 `**' 0.01 `*' 0.05 `.' 0.1 `~' 1 }}}          \\
\end{tabular}

\end{table}

From \Cref{tab:modeling}, we observed that the number of downloads has the most explanatory power with a $\textrm{Wald }\chi^{2}$ value equal to 63.00 when we modeled the probability of \used \npm packages.
The second most important factor in modeling \used packages is the number of stars a package has ($\textrm{Wald }\chi^{2} = 24.21$).
\Cref{fig:nomgram} also shows that \npm packages that have a high number of downloads and received a high number of stars have a high chance to be \used packages.

Our regression analysis also showed that documentation and vulnerability factors have explanatory power as well.
Interestingly, developers reported these two factors in our survey to have a high impact when selecting \npm packages.
With a $\textrm{Wald }\chi^{2}$ value equal to 16.47, packages that have a high percentage of vulnerable versions have higher impact power and the same apply for the size of the readme files with $\textrm{Wald }\chi^{2} = 11.61$.
In addition, \Cref{fig:nomgram} confirms that documentation and vulnerabilities have a positive contribution to the probability of a \npm package being \used.

Interestingly, our regression analysis showed two of the studied factors that have an explanatory power when they are used to model the probability of \used \npm packages, which are the number of badges that the package has and the number of closed issues on Github.
However, our survey results showed that developers tend not to consider these factors when searching for an \npm package to use.
\Cref{tab:modeling} shows that the number of closed issues is the third most important factor with a $\textrm{Wald }\chi^{2}$ value equal to 17.62 while the number of badges is placed fifth, having a value of $\textrm{Wald }\chi^{2}$ equal to 12.21.
Furthermore, \Cref{fig:nomgram} shows the number of badges has a positive contribution with the probability that the package will be \used packages.

In addition, our nomogram analysis shows that the number of contributors as a factor has a negative contribution to the probability of a package being \used. In contrast, the regression analysis showed that this factor has a modest explainable power ($\textrm{Wald }\chi^{2} = 4.41$).

\conclusion{
	In summary, our quantitative analysis complemented developers' perceptions about the factors that they look for when selecting an \npm package to use.
	In particular, our results showed that \used \npm packages tend to possess characteristics that include a high number of downloads, stars, and a higher ratio of closed issues.
	Lastly, in contrast to our qualitative analysis results, our regression analysis showed that a higher number of badges is an essential characteristic of \used \npm packages.
}

\section{Discussion}
\label{sec:survey:discussion}
Our study has many direct benefits for the ecosystem maintainers and the \npm community, particularly package owners and developers who use the \npm packages.
We discussed these implications and benefits in the following.

\textbf{The \npm software ecosystem maintainers should pay attention to certain aspects of the packages when building package search or recommendation tools.}
Several package search tools have been proposed and deployed, which can be classified into two main categories.
The first category based on keyword search~\cite{npm79online,npmDisco64online,npmrank:online}.
These tools are limited since they do not take into consideration the quality aspect of the packages.
Tools from the second category provide package search while considering some quality aspects of the packages, e.g., the \npms tool~\cite{npms13online}.
While \npms is the official search tool used by the official \npm website, it has some limitations.
The main limitation of \npms is that it assigns different weights of the used aspects without a clear justification, which negatively affects the quality of the search engine~\cite{Abdellatif_IST2020}.
Our examination of \npms' source code and documentation shows that \npms arbitrarily gives weights to certain aspects when ranking the packages.
We recommend that the \npm ecosystem could use our results to build more robust search tools.
For example, the \npm maintainers can integrate our rank of the important factors to weigh each factor's contribution when used in a search tool, which they are based on developers' perceptions.

\textbf{Several package characteristics should be carefully examined by developers when choosing an \npm package to depend on in their projects.}
As mentioned earlier, our results indicated that \used packages possess specific characteristics.
For example, our regression analysis results showed that the number of closed issues in the package repository is commonly related to \used packages.
We believe that JavaScript developers can use our results to build systematic guidelines for choosing an \npm package to use.
In fact, there have been several attempts to help developers create such a guideline~\cite{franch2003IEEEsoftware,9f986d4d58online,Semeteys_OSB2008}.
However, their main drawback is that they focused on selecting packages in a specific context or propose general guidelines to select open source components.
In addition, they do not consider package characteristics that \npm provide, such as the number of downloads.

\textbf{To promote their packages, the owner of \npm packages should provide clear indications of their packages' characteristics.}
Gaining more popularity within the software ecosystem requires putting more effort into signalling the published packages' quality.
Overall, all the package factors that our qualitative and quantitative results highlighted are essential factors that package owners can employ to attract more users.
For example, many responses indicated that package documentation is an important factor when looking for a package to use.
Based on these findings, we recommend developers invest more effort in making their package documentation, particularly readme files, clearer and up to date.

\section{Related Work}
\label{sec:survey:related_work}
The increasing trend of depending on software ecosystems by developers has motivated researchers to understand the developer's perspective about using third-party packages.

\citet{Haefliger_2008_CRO} studied the reuse pattern and practices in open source applications.
Their study showed that experienced developers reuse more code than less experienced developers.
Abdalkareem et al.~\cite{Abdalkareem_FSE2017,abdalkareem2017reasons} studied an emerging code reuse practice in the form of lightweight packages in the software ecosystem.
Their study was conducted to understand why developers use trivial packages.
Their results showed that these packages are prevalent in PyPI (Python Package Index), but 70.3\% of the developers considered using these packages is a bad practice.
In addition, \citet{Xu2019} studied the reason behind the reusing and re-implementing of external packages in software applications.
They found that developers often replace their self-implemented methods with external libraries because they were initially unaware of the library or it was unavailable back then.
Later on, when they became aware of a well-maintained and tested package, they replace their own code with that package.
Although developers preferred to reuse code than re-implement it, they replaced an external heavy package with their implementation when they believed that they were only using a small part of its functionalities or if it became deprecated.
\citet{Haenni_2013ACM} conducted a survey with developers about their decision-making while introducing a dependency to their applications.
Surprisingly, the study found that developers generally do not apply rationale while selecting the packages; they used any package that accomplishes the required tasks.

Other work also focused on examining the popularity growth of packages within an ecosystem.
For example, \citet{qiu2018understanding} studied the growth of popular \npm package.
Their finding showed that lifetime, number of dependents, and added new functionalities play significant roles in popularity growth.
\citet{Kyriakos2019free} used network analysis and information retrieval techniques to study the dependencies that co-occur in the development of \npm packages.
Then, they used the constructed network to identify the communities that have been evolved as the main drivers for npm’s exponential growth.
Their findings showed that several clusters of packages can be identified.
\citet{Zerouali_SANER2019}~examined a large number of \npm packages by extracting nine popularity metrics.
They focused on understanding the relationship between the popularity metrics.
They found that the studied popularity metrics were not strongly correlated.

Other work focused on understanding the process used by developers to select packages and attempted to provide some guidelines.
\citet{Pano2018} focused on understanding factors that developers look for when selecting a JavaScript framework (e.g., react).
Based on interviewing 18 decision-makers, they observed a list of factors when choosing a new JavaScript framework, including the community's size behind the framework.
\citet{BiancoIEEESoftware}~provided a list of factors that influence the trustworthiness of open source software components.
Their list had five categories, including quality and economic categories.
Also, in their study, \citet{Hauge2009Workshop} observed that many organizations apply informal selection process based on previous experience, recommendations from experts, and information available on the Internet.
\citet{franch2003IEEEsoftware} investigated to adapt the ISO quality model and assign metrics to be used as a measure for selecting software components.
Their study suggested that relationships between quality entities need to describe explicitly.

The main goal of our study is to examine the characteristics of \used packages within the \npm ecosystem.
In many ways, our study is complementary to prior work since we focus on understanding factors that make a package \used.
That said, our study is one of the only studies to use mixed research methods, which provide us with more complete and synergistic utilization of data than any separate quantitative and qualitative data collection and analysis.

\section{Threats to Validity}
\label{sec:survey:threats_to_validity}
In this section, we discuss the potential threats to the validity of our work.
\subsection{Internal Validity}
Internal validity concerns factors that could have influenced our results. To qualitatively understand the factors that may impact the use of an \npm package, we surveyed JavaScript developers. While we carefully designed our survey based on the guideline provided in~\cite{dillman2011mail}, our survey might have been influenced by some factors. First, our survey participants might poorly understand some of the factor statements. To mitigate this limitation, we conducted a pilot survey where we gave our survey to three expert JavaScript developers and incorporated their feedback about the survey. Second, we had a list of well-defined factors that may impact selecting an \npm package. Even though we choose to study these factors since they are used in the literature and can be easily examined by developers, we may miss some other factors. To mitigate this threat, in our survey, we had one open-ended question, where we asked developers to provide us with any factors that are missed in our survey~\cite{dillman2011mail}. That said, none of our survey responses reported any new factors that can be quantitative.

To recruit participants in our survey, we resorted to developers who publish and use packages from the \npm ecosystem. At the beginning of the survey, we articulated that the purpose of our study is to understand how developers select \npm packages. This description may attract more attention from developers, who use \npm packages more.

\subsection{Construct Validity}
Construct validity considers the relationship between theory and observation in case the measured variables do not measure the actual factors. In our study on the \npm ecosystem, we used \npms platform~\cite{npms13online} to measure various quantitative factors related to download counts, testing, and having a website. Our measurements are only as accurate as \npms; however, given that \npms is the main search tool for npm, we are confident in the \npms metrics. We also used Snyk~\cite{SnykDeve46online} to calculate the number of vulnerabilities that affect the studied packages. Thus, our analyses are as accurate as Snyk dataset. That said, we resorted to using the Snyk data since it has been used by other prior work (e.g.,~\cite{Mahmoud:21:Python,zapata2018towards}). In addition, we wrote a crawler to extract factors from the Github platform through the use of Github API, so our collected data might be affected by the accuracy of this public API. Furthermore, {In our study, we investigated package factors that can be observed in a mechanical way (e.g., examine the Github repository of the package). However, developers might select \npm packages based on a discussion or recommendation by other developers. Thus, our studied factors may not present the whole picture.}

\subsection{External Validity}
Threats to external validity concern the generalization of our findings. In our study, we investigated the factor that impacts \used packages that are published on the \npm ecosystem. Our results might not be generalized to other software ecosystems, such as maven for Java or PyPi for Python. However, since \npm ecosystem is the most popular software ecosystem, this gave us confidence in our results. Also, scientific literature showed that studying individual cases has significantly increased our knowledge in areas such as economics, social sciences, and software engineering~\cite{flyvbjerg2006five}. Second, our dataset that was used in the quantitative analysis presents only open-source packages hosted on GitHub that do not reflect proprietary packages or packages that are hosted on other platforms such as GitLab and BitBucket. Furthermore, we surveyed 118 JavaScript developers, so we do not claim that our results are generalized to other developers who do not know JavaScript or the \npm software ecosystem.

Finally, one criticism of empirical studies results is ``I know it all along'' thought or nothing new is learned%
. However, such common knowledge has rarely been shown to be trusted and is often quoted without scientific and research evidence. Our paper provides such evidence and supports common knowledge (e.g., ``packages with good documentations tend to be \used packages'') while some are challenged (e.g., ``developers do not consider the number of badges when selecting a new package to use.'').

\section{Conclusion}
\label{sec:survey:conclusion}
In this work, we used a mixed qualitative and quantitative approach to investigate the characteristic of \used \npm packages.
We started by identifying seventeen packages selection factors based on our literature review and used by existing online package search tools.
Then, we qualitatively investigated the factors developers look for when choosing an \npm package by surveying 118 JavaScript developers.
Second, we quantitatively examined these factors by building a logistic regression model using a dataset of {2,527} \npm packages divided into \used and \unused packages.

Among our main findings, we highlighted that JavaScript developers believe that \used packages are well-document, receive a high number of stars on GitHub, have a large number of downloads, and do not suffer from security vulnerabilities.
Moreover, our regression analysis complemented what developers believe about \used packages and showed the divergences between the developers' perceptions and the characteristics of \used packages.

\section{Acknowledgments}
The authors would like to thank all the developers and managers who provided their valuable feedback during the survey.

\bibliographystyle{cas-model2-names}
\bibliography{bibliography}

\end{document}